\documentclass[twocolumn]{aastex631}

\usepackage{graphicx}	
\usepackage{amsmath}	
\usepackage{amssymb}	
\usepackage{xcolor}
\usepackage{multirow}
\usepackage{soul}

\def\omA{\omega_{\rm A}}

\newcommand{\Nth}{N_{\rm th}}
\newcommand{\Nmu}{N_{\mu}}
\newcommand{\kapth}{\kappa_{\rm th}}
\newcommand{\kapmu}{\kappa_{\mu}}
\newcommand{\om}{2\Omega}

\newcommand{\fmw}{\frac{\omA^2}{\om}}
\def\omAeta{\omega_{\rm A}^{\eta,\min}}
\def\omAnu{\omega_{\rm A}^{\nu,\min}}

\def\bk{\boldsymbol{k}}

\def\kTI{k_{\rm TI}}
\def\tw{t_q}

\def\SB{SB24}

\def\Eq{Equation}
\def\Eqs{Equations}
\def\beq{\begin{equation}}
\def\eeq{\end{equation}}

\begin{document}

\title{ Zones of Tayler Instability in Stars}

\author{Valentin A. Skoutnev}
\affiliation{Physics Department and Columbia Astrophysics Laboratory, Columbia University, 538 West 120th Street New York, NY 10027,USA}
\affiliation{Max Planck Institute for Astrophysics, Karl-Schwarzschild-Str. 1, D-85741, Garching, Germany}

\author{Andrei M. Beloborodov}
\affiliation{Physics Department and Columbia Astrophysics Laboratory, Columbia University, 538 West 120th Street New York, NY 10027,USA}
\affiliation{Max Planck Institute for Astrophysics, Karl-Schwarzschild-Str. 1, D-85741, Garching, Germany}



\begin{abstract}

The Tayler instability (TI) of toroidal magnetic fields is a candidate mechanism for driving turbulence, angular momentum (AM) transport, and dynamo action in stellar radiative zones. Recently \cite{Skoutnev_2024} revisited the linear stability analysis of a toroidal magnetic field in a rotating and stably stratified fluid. In this paper, we extend the analysis to include both thermal and compositional stratification, allowing for general application to stars. We formulate an analytical instability criterion for use as a ``toggle switch" in stellar evolution codes. It determines when and where in a star the TI develops with a canonical growth rate as assumed in existing prescriptions for AM transport based on Tayler-Spruit dynamo. We implement such a ``toggle switch" in the MESA stellar evolution code and map out the stability of each mode of the TI on a grid of stellar evolution models. In evolved lower mass stars, the TI becomes suppressed in the compositionally stratified layer around the hydrogen burning shell. In higher mass stars, the TI can be active throughout their radiative zones, but at different wavenumbers than previously expected. 

\end{abstract}



\section{Introduction}

AM transport in the radiative zones of stars remains an important problem in stellar physics. Evolving stars experience structural adjustments and surface torques from winds that lead to differential rotation of their interiors \citep{maeder2000evolution,maeder2008physics}. Without redistribution of AM, the compact cores of evolved stars would rotate orders of magnitude faster than their envelopes and leave behind rapidly rotating stellar remnants. By contrast, observations show relatively slow internal stellar rotation rates \citep{beck2012fast,mosser2012spin,deheuvels2012seismic,di2016internal,gehan2018core,tayar2019core,kuszlewicz2023mixed,li2024asteroseismic,mosser2024locked}  and small initial spins of stellar remnants \citep{heger2005presupernova,suijs2008white,kawaler2014rotation,hermes2017white}. This broadly suggests efficient transport of AM in stellar interiors.

The transport mechanism remains poorly understood (for a review, see \cite{aerts2019angular}). Note that it has to be sustained in a broad range of radii without interruption; blocking it even in a narrow layer at some radius would isolate the core AM, leaving the core with fast rotation. One possibility is turbulent transport,\footnote{Alternative processes include internal gravity waves from nearby convection zones \citep{fuller2014angular,blouin20233d} or large scale magnetic fields deposited from earlier stages of evolution \citep{kissin2015rotation,takahashi2021modeling}.} which operates in the presence of instabilities. Hydrodynamic instabilities are typically inefficient and often inhibited in regions of strong compositional stratification near the edge of evolving stellar cores \citep{heger2000presupernova}. More efficient turbulent transport may occur in the presence of magnetohydrodynamic (MHD) instabilities. 

In particular, the TI of toroidal magnetic fields \citep{tayler1973adiabatic,spruit1999differential} is a promising candidate because it develops in a stably stratified fluid more easily than other MHD instabilities (e.g. magnetic buoyancy \citep{acheson1979instability,hughes1985magnetic,spruit1999differential} and the magnetorotational instability \citep{wheeler2015role,jouve2020interplay}). Differential rotation may naturally produce magnetic configurations prone to TI as it winds any existing, weak, radial magnetic field $B_R$ into a much stronger toroidal magnetic field $B_\phi$. The instability is active in the polar regions of a star and the generated turbulence may support a 
dynamo loop. This scenario, known as the Tayler-Spruit dynamo \citep{spruit2002dynamo}, seems capable of explaining AM transport, although its efficiency depends on the debated saturation level of the TI and unclear statistical properties of the excited turbulence \citep{braithwaite2006differential,zahn2007magnetic,arlt2011amplification,guerrero2019global,ji2023magnetohydrodynamic,monteiro2023global,petitdemange2023spin,barrere2023numerical}.
Despite several uncertainties, AM transport enabled by the TI is widely invoked as a leading explanation of the rotation rates measured in stellar interiors \citep{heger2005presupernova,cantiello2014angular,braithwaite2017magnetic,aerts2019angular,ma2019angular,eggenberger2022rotation,schurmann2022spins,rosales2024double}.

The linear stability analysis of the TI was recently revisited in \cite{Skoutnev_2024} (\SB). We systematically examined each wave branch of the dispersion relation, which led to discovery of new unstable modes and revision of previously known modes. Our analysis also revealed the physical picture of the TI, in particular how the instability of large-scale toroidal fields in rotating stars is enabled by microphysical diffusivities.
While strong Coriolis forces hinder the TI, diffusive processes promote instability on length scales where diffusive and Coriolis timescales are comparable, allowing magnetic loops to rearrange and release magnetic energy. The TI can be enabled by the diffusion of the fluid momentum, magnetic field, temperature, and composition. The corresponding diffusivities will be denoted as $\nu$, $\eta$, $\kapth$, and $\kapmu$, respectively.\footnote{The diffusivities are determined by the local composition, temperature, and density. Generally, the thermal diffusivity (mediated by photons) is the largest, followed by the viscosity (mediated by ions and photons), and then by the compositional diffusivity (mediated by ions), so $\kapth>\nu>\kapmu$. The ratios $Pm=\nu/\eta$ and $Cm=\kapmu/\eta$  can be smaller or larger than unity.} \SB~ extended the TI analysis to fluids with any magnetic Prandtl number $Pm=\nu/\eta$, including $Pm\gg 1$. The latter turns out to be the relevant limit for stars significantly more massive than the Sun, as will be shown in the present paper. 

In this paper, we complete the general analysis of TI in stars. First, we extend the results of \SB~to include both thermal and compositional stratification (only one type of stratification was considered in \SB) and summarize the instability criteria for each wave branch. The criteria are obtained by analytically solving for the growth rates and confirming with numerical solutions. The systematic analysis allows us to overcome some limitations of previous works. In particular, \cite{spruit1999differential} focused on stars with mass $M\lesssim 1 M_\odot$ where the magnetic diffusivity dominates over the viscous and compositional diffusivities. This limit is inapplicable in higher mass stars and in evolved low mass stars, where diffusivities vary by orders of magnitude across the wide range of temperatures and densities \citep{jermyn2022atlas}. Furthermore, the TI was previously treated with a heuristic approach based on a marginal stability calculation. It did not correctly distinguish the different wave branches of instability and, in some cases, led to incorrect identification of the wavenumbers of the most unstable modes. We also find that previous treatment of thermal+compositional stratification using an effective Brunt-V\"ais\"al\"a frequency \citep{spruit1999differential,spruit2002dynamo} is deficient. 

After formulating the revised stability criteria (Section~\ref{sec:LinearAnalysisResults}),  we implement them in the MESA stellar evolution code \citep{paxton2010modules,paxton2013modules,paxton2015modules,paxton2018modules,paxton2019modules,jermyn2023modules} and examine the onset of TI in stellar interiors (Section~\ref{sec:Stars}). The presence or absence of instability is of particular interest for the core-envelope transition in evolved stars. These transitional layers can act as a barrier for AM transport because of strong compositional stratification left behind by nuclear shell burning. We find that TI remains unimpeded throughout stellar evolution only in sufficiently massive stars, provided that the radial component of their magnetic fields is sufficiently weak. In evolved low-mass stars, we find that the TI is suppressed by strong compositional gradients, contrary to previous expectations. 

\section{Tayler Instability with Thermal and Compositional Stratification}
\label{sec:LinearAnalysisResults} 
We are interested in the TI of toroidal magnetic fields, $B_\phi$, in the rotating and stably stratified (radiative) zones of stellar interiors. Stable stratification can have contributions from both thermal and compositional gradients, which are associated with their own Brunt-V\"ais\"al\"a frequencies $\Nth$ and $\Nmu$, and diffusivities $\kapth$ and $\kapmu$ (for a review, see \cite{garaud2018double}). 

Stratification in stars is typically strong, with $\max\{\Nth,\Nmu\}\gg\Omega$, where $\Omega$ is the rotation rate. Due to efficient horizontal transport of AM in radiative zones, rotation is approximately constant on spherical shells $\Omega(R,\theta)\approx \Omega(R)$ (where $\{R,\theta,\phi\}$ are spherical coordinates) \citep{zahn1992circulation}. Evolution of the star causes the build up of differential rotation. The radial shear then generates $B_\phi$ through the winding of an initially embedded weak radial field $B_R$,
\begin{align}
    \label{eq:winding}
    \partial_t B_\phi=B_R\, q\Omega \sin\theta,\qquad 
    q=\frac{d\ln\Omega}{d\ln R}.
\end{align}
For a magnetic field that sources its energy from the differential rotation, the upper bound on its energy density is $B_\phi^2/8\pi\sim \rho r^2\Omega^2/2$, where $\rho$ is the fluid density and $r=R\sin\theta$ is the cylindrical radius. This implies an Alfv\'en frequency $\omA= B_\phi/\sqrt{4\pi\rho r^2}$ smaller than $\Omega$. Furthermore, $\omA\ll\Omega$ is normally satisfied in the models of AM transport employing a Tayler-Spruit dynamo. Therefore, we will investigate the TI in stars that satisfy
\beq
  \omA\ll\Omega\ll\max\{\Nth,\Nmu\}.
\eeq

A sufficiently wound up toroidal field may reach a threshold for instability, driving turbulence and local transport of AM. To obtain the linear instability criteria one must make a choice for the magnetic field configuration, which is generally not known. The  simplest assumption is that $B_\phi$ results from winding of a dipole field ($B_R\propto \cos\theta$), which generates the configuration
\beq\label{eq:Bphi_config}
    B_\phi\propto \sin\theta\cos\theta.
\eeq
Near the polar axis of the star, $|\cos\theta|\approx 1$, this configuration can be described in a cylindrical geometry as $B_\phi\propto r$ \citep{spruit1999differential,zahn2007magnetic}. In fact, near the rotation axis (where TI develops), Stokes theorem requires a similar profile of $B_\phi$ assuming any finite current density along the axis. Hereafter, we consider the stability of toroidal magnetic field configurations given by \Eq~(\ref{eq:Bphi_config}).\footnote{Toroidal fields with stronger gradients with respect to $r$ may also, in principle, exist. Such configurations are more prone to instability; their analysis is presented in Appendix \ref{ap:Large_p} for completeness.}

In a non-rotating star, the magnetic configuration would be unstable on the Alfv\'enic timescale $\omA^{-1}$ \citep{tayler1973adiabatic}. Fast rotation tends to stabilize it. In particular, if the star is treated as an ideal MHD fluid, the instability disappears when $\Omega\gg\omA$ \citep{pitts1985adiabatic,spruit1999differential} as
Coriolis forces act on short rotation timescales $ t_{\Omega}=(2\Omega)^{-1}$ to prevent the growth of seed motions of magnetic field loops. However, diffusive processes can break rotational constraints and enable the TI. As shown in \SB, each diffusive process can independently lead to instability. 

While it may seem counter-intuitive that the stability of a large-scale magnetic field depends on microphysical diffusivities, note that instability is enabled for modes with large (and nearly radial) wavevectors $\bk$. These modes are allowed by stratification because they involve nearly horizontal displacements $\boldsymbol{\xi}$ (as required by $\bk\cdot\boldsymbol\xi=0$ for approximately incompressible perturbations), avoiding the large potential energy cost of radial displacements. It is across the short radial scales $k_R^{-1}\approx k^{-1}$ that diffusive processes are able to operate sufficiently fast to compete with Coriolis forces. Each diffusive process has a timescale:
\beq
    t_\eta=\frac{1}{\eta k^2},\; t_\nu=\frac{1}{\nu k^2},\; t_{\kapth}=\frac{\kapth k^4}{k_\theta^2 \Nth^2},\; t_{\kapmu}=\frac{\kapmu k^4}{k_\theta^2 \Nmu^2},
\eeq
where $k_\theta\ll k$ is the latitudinal component of the wavevector. \SB~found that instability peaks at wavenumbers $k=\kTI$ where the rotation timescale  $t_\Omega$ is comparable to a diffusion timescale. Equating $t_\Omega$ with $t_\eta$, $t_\nu$, $t_{\kapth}$, and $t_{\kapmu}$, one finds the four canonical wavenumbers 
\begin{align}
    &k_\eta=\left(\frac{\om}{\eta}\right)^{1/2}, \qquad k_\nu=\left(\frac{\om}{\nu}\right)^{1/2},\;\\
    &k_{\kapth}=\left(\frac{k_\theta^2\Nth^2}{\om\kapth }\right)^{1/4}, \qquad k_{\kapmu}=\left(\frac{k_\theta^2\Nmu^2}{\om\kapmu }\right)^{1/4}.
\end{align}

The four wavenumbers $\kTI$ are associated with two classes of waves that are supported in a rotating, magnetized fluid: inertial waves (IW) and magnetostrophic waves (MW). The TI is an overstability of the waves, described by a complex frequency $\omega=\omega_r+i\gamma$ with $\gamma>0$. Instability of IW is enabled by magnetic diffusivity at $\kTI=k_\eta$; these waves oscillate quickly, with a real frequency $|\omega_r|\sim \om\gg\omA$. Instability of MW is enabled by viscosity at $k_\nu$ and by thermal or compositional diffusion at $k_{\kapth}$ or $k_{\kapmu}$; these waves have a low $|\omega_r|\sim \omA^2/\om\ll\omA$.

The maximum growth rate $\gamma$ that can occur at each $\kTI$ is around $\gamma^{\max}= \omA^2/4\Omega$. It is much smaller than the growth rate $\gamma_{\Omega=0}\approx\omA$ that would occur in a non-rotating star. The reduction of  $\gamma^{\max}$ in the fast rotation regime by the factor of $\omA/4\Omega$ is explained in SB24.
 
Below we formulate the instability criteria for each $\kTI$ and then apply them to track the presence of TI modes throughout the interior of evolving stars.

\subsection{Instability Criteria}
\label{sec:InstCrit}

\begin{table*}
     \centering
    \begin{tabular}{c c c c }
         Wave branch & Peak wavenumber & Sub-case & Interval of $\omA$ that gives instability with $\gamma^{\max}= \displaystyle \frac{\omA^2}{4\Omega}$ \\[1.5ex]
        \hline\hline
        \\[0.025ex]
         \multirow[b]{2}{*}{Inertial}  & \multirow[b]{2}{*}{$\displaystyle {k_\eta=\left(\frac{\om}{\eta}\right)^{1/2}}$} & $\displaystyle{\sum_i\frac{N_i^2}{4\Omega^2}\left(1+\frac{\kappa_i^2}{\eta^2}\right)^{-1}<\left(\frac{\eta k_\theta^2}{\om}\right)^{-1}}$ & $\displaystyle{ \frac{\omA^2}{4\Omega^2}> \max\left\{Pm,\sum_i \frac{N_i^2}{4\Omega^2}\left(\frac{\kappa_i k_\theta^2}{2\Omega}\right)\left(1+\frac{\kappa_i^2}{\eta^2}\right)^{-1}\right\}}$  \\ 
         &  & $\displaystyle{\sum_i\frac{N_i^2}{4\Omega^2}\left(1+\frac{\kappa_i^2}{\eta^2}\right)^{-1}>\left(\frac{\eta k_\theta^2}{\om}\right)^{-1}}$& No instability\\
         \hline 
         \\[0.025ex]
         Magnetostrophic& $\displaystyle{k_\nu=\left(\frac{\om}{\nu}\right)^{1/2} }$ & $-$&   $\displaystyle{\frac{\omA}{\om}>\max\left\{\sum_i\Theta\left(\frac{k_{\kappa_i}}{k_\nu}-1\right)\frac{N_i}{2\Omega}\left(\frac{\nu k_\theta^2}{\om}\right)^{1/2}, Pm^{-1/2}\right\} }$ 
    \end{tabular}
    \caption{ 
    Summary of instability criteria for the TI modes mediated by magnetic ($\eta$) or viscous ($\nu$) diffusion in a rotating star with $\omA\ll\om$. The maximum growth rate $\gamma^{\max} = \omA^2/4\Omega$ is attained in the wave branch shown in the first column at the canonical wavenumber shown in the second column; the instability
    conditions are stated in the third and fourth columns. 
    }
    \label{tab:InstabilityCriteria_etaandnu}
\end{table*}

\begin{table*}
     \centering
    \begin{tabular}{c c c}
        Peak wavenumber & Sub-case & Interval of $\omA$ that gives instability with $\gamma^{\max}= \displaystyle \frac{\omA^2}{4\Omega}$ \\[1.5ex]
         \hline\hline
         \\[0.025ex]
        \multirow[b]{2}{*}{$\displaystyle{k_{\kapmu}=\left(\frac{k_\theta^2 \Nmu^2}{\kapmu \om}\right)^{1/4}}$}  & $\displaystyle{\Nmu>\Nth\left(\frac{\kapmu}{\kapth}\right)^{1/2}}$ &  $\displaystyle{\left(\frac{\eta}{\kapmu}\right)^{1/2}\left(\frac{\Nmu }{\om}\right)^{1/2}\left(\frac{\kapmu k_\theta^2}{\om}\right)^{1/4}<\frac{\omA}{\om}<\left(\frac{\Nmu }{\om}\right)^{1/2}\left(\frac{\kapmu k_\theta^2}{\om}\right)^{1/4}}$  \\
        & $\displaystyle{\Nmu<\Nth\left(\frac{\kapmu}{\kapth}\right)^{1/2}}$ &  No instability    \\[2.0ex]
        \hline
        \\[0.025ex]
        \multirow[b]{3}{*}{$\displaystyle{k_{\kapth}=\left(\frac{k_\theta^2 \Nth^2}{\kapth \om}\right)^{1/4}}$} &$\displaystyle{\Nmu<\Nth\left(\frac{\kapmu}{\kapth}\right)^{1/2}}$ &  $\displaystyle{\left(\frac{\eta}{\kapth}\right)^{1/2}\left(\frac{ \Nth}{\om}\right)^{1/2}\left(\frac{ \kapth k_\theta^2}{\om}\right)^{1/4}}<\frac{\omA}{\om}<\left(\frac{ \Nth}{\om}\right)^{1/2}\left(\frac{ \kapth k_\theta^2}{\om}\right)^{1/4}$ 
        \\
        & $\displaystyle{\Nth\left(\frac{\kapmu}{\kapth}\right)^{1/2}<\Nmu<\Nth}$  &  $\displaystyle{\max\left\{\left(\frac{\eta}{\kapth}\right)^{1/2},\frac{\Nmu}{\Nth}\right\}\left(\frac{ \Nth}{\om}\right)^{1/2}\left(\frac{ \kapth k_\theta^2}{\om}\right)^{1/4}}<\frac{\omA}{\om}<\left(\frac{ \Nth}{\om}\right)^{1/2}\left(\frac{ \kapth k_\theta^2}{\om}\right)^{1/4}$\\
        & $\displaystyle{\Nth<\Nmu}$  &  No instability
    \end{tabular}
    \caption{ Summary of the instability criteria for the TI modes mediated by compositional ($\kapmu$) or thermal ($\kapth$) diffusivity, which are excited in the MW branch. The columns are arranged similar to Table~\ref{tab:InstabilityCriteria_etaandnu}. The stated criteria assume that thermal diffusivity exceeds the compositional diffusivity, $\kapth>\kapmu$, and that viscous effects are negligible. The latter condition is equivalent to $k_\nu\gg\kTI$ (where $\kTI=k_{\kappa_{\rm th}}$ or $k_{\kappa_\mu}$); otherwise, the instability peak is suppressed by viscous effects.
    }
    \label{tab:InstabilityCriteria_thandmu}
\end{table*}

The TI is not hindered when instability exists with the maximum growth rate $\gamma^{\max}$ for at least one of the four canonical wavenumbers $\kTI$. The instability depends on the local fluid parameters (the plasma diffusivities, rotation rate, and type of stratification) and the magnetic field strength $B_\phi$ or, equivalently, the Alfv\'en frequency $\omA$. We present the analytic derivations of the instability criteria for each $\kTI$ in Appendices  \ref{ap:LIA}, \ref{ap:IWs}, \ref{ap:MWs}, and display the results in Tables \ref{tab:InstabilityCriteria_etaandnu} and \ref{tab:InstabilityCriteria_thandmu}. This extends the results of \SB~to stars where both types of stratification are present, $N_{\rm th}\neq 0$ and $N_\mu\neq 0$.\footnote{In the case of a single type of stratification, Tables~1 and 2 are reduced to Table~1 in \SB, e.g. by setting $\Nmu=0$.} 
Below we summarize the main features of the TI.

The key parameter for TI enabled by viscous or magnetic diffusion is the magnetic Prandtl number:\footnote{Strong stratification may stabilize the TI at $k_\eta$ and $k_\nu$, see Table~\ref{tab:InstabilityCriteria_etaandnu} for details.}
\begin{itemize}
    \item IW at $k_\eta$ are unstable where $Pm\ll1$.
    \item MW at $k_\nu$ are unstable where $Pm\gg1$.
\end{itemize}
For TI enabled by buoyancy effects, an important factor is the relative strength of compositional and thermal stratification, $\Nmu/\Nth$. In the regime relevant to stars, $\kapth>\kapmu$,\footnote{
In the opposite case of $\kapth<\kapmu$, the roles of $\Nth$ and $\Nmu$ in the instability criteria would be swapped, as follows from the symmetry of the dispersion relation under $\Nth\leftrightarrow\Nmu$.
} 
we find
\begin{itemize}
    \item MW at $k_{\kapth}$ are unstable where thermal stratification is dominant, $\Nth\gg\Nmu$, and thermal diffusion is faster than magnetic diffusion, $\kapth\gg\eta$.
    
    \item MW at $k_{\kapmu}$ are unstable where compositional stratification is sufficiently strong, $\Nmu\gg\Nth\sqrt{\kapmu/\kapth}$, and compositional diffusion is faster than magnetic diffusion, $\kapmu\gg\eta$. 
\end{itemize}
To highlight the importance of the ratio $\kapmu/\eta$ for TI in a compositionally stratified fluid, we define the dimensionless parameter 
\begin{equation}
 Cm\equiv\frac{\kapmu}{\eta}.
\end{equation}
The instability criteria in stars (where generally $\kapth\gg \nu,\kapmu,\eta$) are sensitive to three dimensionless parameters:  $\Nmu/\Nth$, $Pm$, and $Cm$. These parameters can vary sharply near nuclear burning regions (e.g. at the core boundary in evolved stars). 

\begin{figure*}
    \centering
    \includegraphics[width=\linewidth]{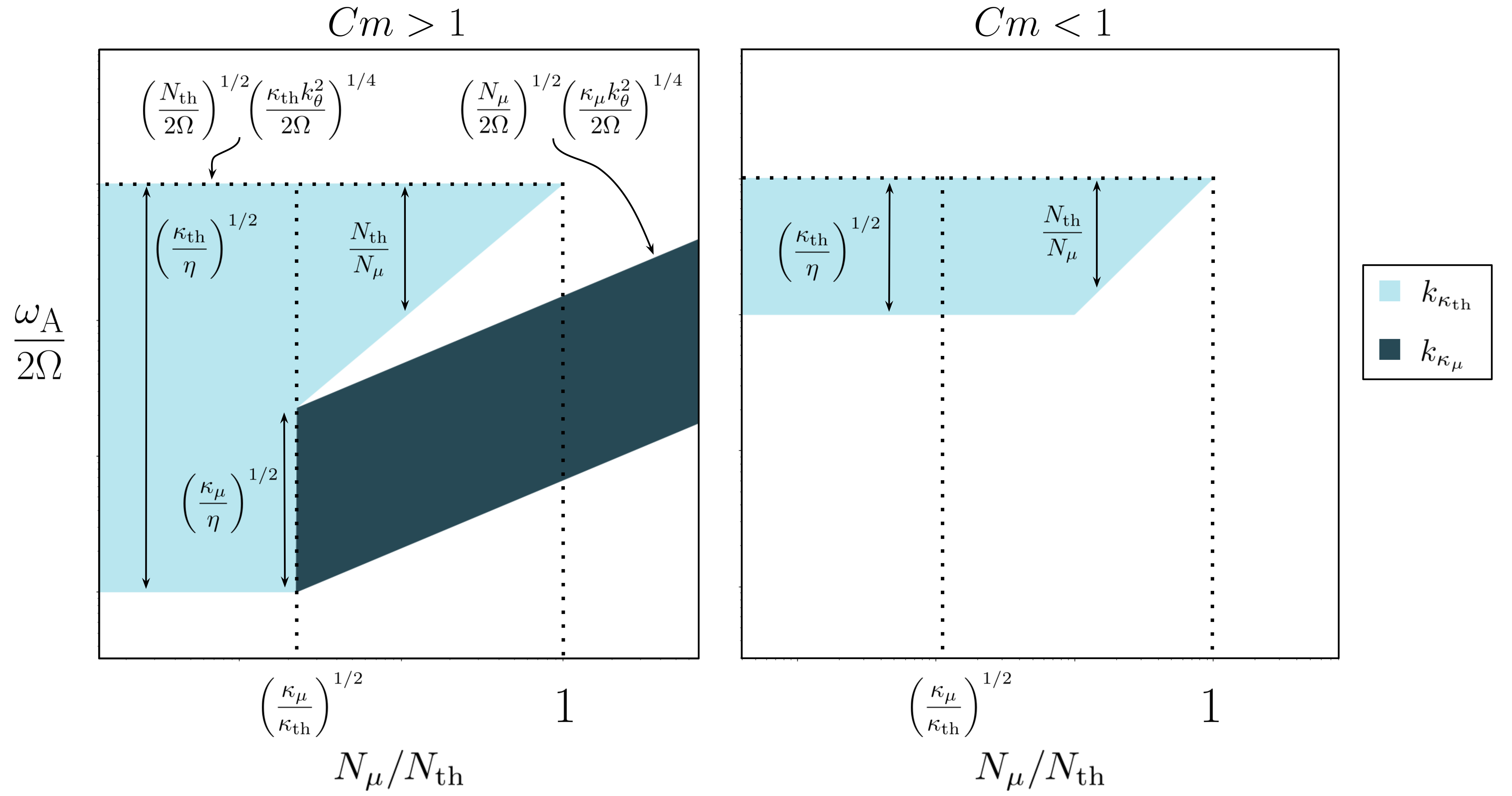}
    \caption{Intervals of $\omA/\om$ that give the MW instability peak at $\kTI=k_{\kapth}$ (light blue) or $\kTI=k_{\kapmu}$ (dark blue) vs. $\Nmu/\Nth$. The two limits of the key parameter $Cm=\kapmu/\eta$, $Cm>1$ and $Cm<1$, are shown on the left and right, respectively. The figure assumes the typical regime for stars: $\kapth\gg\kapmu,\eta$ (dominant thermal diffusivity) and $k_\nu\gg k_{\kapth},k_{\kapmu}$ (weak effects of viscosity). The axes are on logarithmic scales.
    }
    \label{fig:TwoStrat_Stab}
\end{figure*}

The TI enabled by viscous or magnetic diffusion (at $k_\nu$ or $k_\eta$) occurs once $\omA$ exceeds a threshold (Table~\ref{tab:InstabilityCriteria_etaandnu}). By contrast, the TI enabled by thermal or compositional diffusion (at $k_{\kapth}$ or $k_{\kapmu}$) exists in a finite interval of $\omA$ (Table~\ref{tab:InstabilityCriteria_thandmu}). The lower and upper limits on $\omA$ are determined by physical processes that are easiest to understand when a single type of stratification is dominant, e.g. thermal. The toroidal magnetic field must be strong enough for the growth rate at $k_{\kapth}$ to exceed the rate of suppression by magnetic diffusion: $\omA^2/4\Omega>\eta k_{\kapth}^2$. On the other hand, it must be weak enough to satisfy $\omA^2/\om<\kapth k_{\Nth}^2$, so that the MW oscillation is slower than buoyancy diffusion at wavenumber $k_{\Nth}=k_\theta \Nth/\omA$ (not $k_{\kapth}$, see \SB~for details). Note also that the intervals of $\omA$ for which the $k_{\kapth}$ and $k_{\kapmu}$ modes are unstable both scale with the free parameter $k_\theta^{1/2}$, which has a minimum value $k_\theta\sim 1/R$. A larger $k_\theta$ shifts the instability intervals to larger $\omA$.

When both compositional and thermal stratification are present, their buoyancy responses can interact and the TI depends on the ratio $\Nmu/\Nth$ (see Appendix \ref{ap:MW_buoyancy} for details). Figure~\ref{fig:TwoStrat_Stab} shows the unstable intervals of $\omA$ for $k_{\kapth}$ and $k_{\kapmu}$ as a function of $\Nmu/\Nth$. Note that the TI cannot occur simultaneously at $k_{\kapth}$ and $k_{\kapmu}$.

\subsection{Comparison with Previous Works}

Previous works missed the instability peaks at $k_\eta$ and $k_\nu$. Furthermore, our results for the instability at $k_{\kapth}$ and $k_{\kapmu}$ differ from previous results. The disagreement stems from the common but inappropriate use of an effective Brunt-V\"ais\"al\"a frequency. For a single type of stratification (e.g. thermal), it is defined as follows
\begin{eqnarray}
\label{eq:Neff}
    N_{\rm eff}^2\equiv\frac{\Nth^2}{1+\kapth k_{N_{\rm eff}}^2\Omega/\omA^2},\quad k_{N_{\rm eff}}=k_\theta\frac{ N_{\rm eff}}{\omA}.
\end{eqnarray} 

Previous studies used $N_{\rm eff}$ to identify the characteristic unstable wavenumber of the MW branch as $k_{N_{\rm eff}}$ and stated the condition for instability as $\omA^2/\om>\eta k_{N_{\rm eff}}^2$ (e.g. \cite{spruit1999differential,spruit2002dynamo}). This approach agrees with our results only when buoyancy diffusion is fast, $\kapth k_{\Nth}^2\gg\omA^2/\om$, so that $k_{N_{\rm eff}}=k_{\kapth}$. This occurs for $\omA$ in the interval
\begin{align}
    \label{eq:omega_th_interval}
    &\left(\frac{\eta}{\kapth}\right)^{1/2}\omega_{\rm th}<\omA<\omega_{\rm th},\\
    \label{eq:omega_th}
    \omega_{\rm th}&\equiv2\Omega\left(\frac{\Nth}{2\Omega}\right)^{1/2}\left(\frac{\kapth k_\theta^2}{2\Omega}\right)^{1/4}.
\end{align}
However,  when buoyancy diffusion is slow, which occurs for $\omA>\omega_{\rm th}$, the MW branch is stable. Instead, the fifth branch of the TI (which degenerates to the $\omega=0$ mode in the ideal MHD limit \citep{zahn2007magnetic}) can be unstable at $k_{N_{\rm eff}}=k_{\Nth}$. Previous works using marginal stability analysis misclassified instability at $k_{\Nth}$ in the small $\kapth$ limit as a mode of the MW branch, instead of the fifth branch. \SB~showed that the fifth branch is unstable with a maximum growth rate $\gamma_5^{\max}\sim (\eta k_\theta^2 N^2/16\Omega)^{1/2}$ which is independent of $\omA$ and comparable to $\sim\omA^2/4\Omega$ only for a narrow interval of $\omA^2/2\Omega\sim \eta k_N^2$ (where buoyancy and magnetic diffusivity can interact); otherwise, $\gamma_5^{\max}$ is relatively small. Therefore, in the present paper, we examine the fifth branch only when all four canonical modes are stable.

Considering the case of a single stratification type (e.g. thermal as discussed above) is sufficient to see the issue with using $N_{\rm eff}$. An effective Brunt-V\"ais\"al\"a frequency can also be defined in the  general case with both thermal and compositional stratification \citep{spruit2002dynamo}, and its use leads to incorrect conclusions for similar reasons.

\subsection{Role of the Radial Field and Differential Rotation}
\label{sec:Randq}
We have so far examined an idealized setup of the TI for a purely toroidal magnetic field, neglecting the radial component of the magnetic field $B_R$ and the differential rotation, $q=d\ln\Omega/d\ln R$. However, finite values of both $B_R\neq0$ and $q\neq 0$ are needed to generate a toroidal field in the first place (\Eq~\ref{eq:winding}). Below we discuss the conditions for their effects on the TI to be small.

\paragraph {Radial field} 
In the presence of a radial field, magnetic tension forces oppose horizontal fluid motions with large radial shears that are characteristic of the TI. As a result, a sufficiently strong $B_R$ can suppress the TI \citep{braithwaite2009axisymmetric}. The condition for $B_R$ to be negligible for a mode with wavenumber $\kTI$ is 
\begin{equation}
\label{eq:radial_field}
    \omA\gg (\kTI R)\omega_{\rm A}^R,
\end{equation}
where $\omega_{\rm A}^R=B_R/\sqrt{4\pi\rho r^2}$ is the Alfv\'en frequency of the radial magnetic field. We assume in this work that the local $B_R$ in the star is sufficiently weak that the condition in \Eq~(\ref{eq:radial_field}) is satisfied.

\paragraph{Differential rotation} Linear stability analysis with $q=0$ remains valid for finite values of $q$ below some threshold. At the threshold, effects of winding become significant on the length and time scales of the TI modes. Above the threshold (which is different for each mode of the TI), the linear stability analysis does not apply and this regime requires further study outside the scope of this paper. Below we state a simple estimate for the threshold $q$ below which differential rotation can be neglected. A more formal analysis using the dispersion relation with finite $q$ is given in Appendix \ref{ap:q}.

A radial magnetic field perturbation $b_R$ driven by the TI (with $q=0$) is coherent on the radial scale $\Delta R\sim \pi/\kTI$ and the timescale $\sim |\omega_r|^{-1}$. Differential rotation shears the coherent patch of radial field on the timescale 
\beq
 \tw\sim \frac{2\pi}{\Delta R\, |d\Omega/dR|} \sim \frac{2\kTI R}{|q|\Omega}. 
\eeq
The shear distortion may be insignificant only if $\tw\gg |\omega_r|^{-1}$, so that $b_R$ can oscillate many times before the mode is substantially sheared. This condition can be written as an upper bound on the differential rotation 
\beq
\label{eq:q_condition}
|q|\ll \kTI R\frac{|\omega_r|}{\Omega}.
\eeq

For IW,
$|\omega_r|\approx 2\Omega$. Then, the condition $\tw\gg |2\Omega|^{-1}$ implies that the linear stability analysis with $q=0$ may be applicable when 
\beq
    |q|\ll k_\eta R \qquad \mathrm{(TI\;via\;IW).}
\eeq
This condition is satisfied with typical parameters $k_\eta R\gg1$ and $q\sim1$. 

For MW, the condition $\tw\gg |\omA^2/2\Omega|^{-1}$ involves the magnetic field strength $B_\phi\propto\omA$. Since MW oscillate slower for weaker fields,
differential rotation will distort a mode before a single oscillation is completed if $\omA$ is too small. For negligible distortion, $\omA$ must satisfy
\beq
\label{eq:q_condition_MW}
    \omA\gg\omega_{\rm A,q}\equiv \om\left(\frac{q}{\kTI R}\right)^{1/2} \quad \mathrm{(TI\;via\;MW)},
\eeq
where $\kTI=k_{\kapth}$, $k_{\kapmu}$, or $k_{\nu}$. This condition is not trivially satisfied. Therefore, MW instability calculated with $q=0$ may be justified only in regions where differential rotation is sufficiently weak and the expected dynamo-saturated toroidal field strengths are above the threshold given in \Eq~(\ref{eq:q_condition_MW}).

\section{Application to stellar models}
\label{sec:Stars}

In this Section, we examine the properties of the TI across a stellar interior for a few representative stellar masses and evolutionary phases. We map out the regions unstable to the four canonical TI modes and determine the easiest mode to destabilize. We first focus on a fiducial $1.5M_\odot$ star in detail, which we find is representative of low mass stars $\lesssim 4M_\odot$. We then examine higher mass stars separately because they have substantively different profiles of $Pm$ and $Cm$ during their evolution, to which the instability criteria are highly sensitive.  

\subsection{Stellar models}

Stellar models are computed using the MESA stellar evolution code. They are evolved from zero age main sequence (ZAMS) with an initially uniform rotation profile and solar metallicity $Z=0.02$. Standard parameters are used for hydrodynamic mixing processes, the convective overshoot (`step'), and mass loss prescriptions (`Dutch' with efficiency $\eta=0.5$ for more massive stars $M>3M_\odot$). However, prescriptions for AM transport due to the Tayler-Spruit dynamo are turned off, as we only seek to characterize the stability of TI modes. We do not expect the regions of stability to change if the AM transport was self-consistently included because the main parameters influencing the TI ($\Nmu/\Nth$, $Pm$, and $Cm$) are nearly independent of the rotational profile. Rotational mixing by the TI can modify the profile of $\Nmu$, but we have found the effect to be negligible for our purposes. In all cases, the mass coordinate and time step resolution have been increased until models are reasonably converged.

To compute the stability of TI modes in MESA, we implement the logic and analytical expressions presented in Table \ref{tab:InstabilityCriteria_etaandnu} and \ref{tab:InstabilityCriteria_thandmu}. Each $\kTI$ has a minimum and maximum Alfv\'en frequency associated with the interval of toroidal field strengths for which it is unstable,
\beq
\omA^{\rm TI,\min}<\omA<\omA^{\rm TI,\max}.
\eeq
The $k_\eta$ and $k_\nu$ modes are unstable for $\omA$ above a minimum that we denote as $\omAeta$ and $\omAnu$, respectively. On the other hand, the $k_{\kapth}$ and $k_{\kapmu}$ modes can be unstable within an interval of $\omA$, which we denote as $\omA^{\kapth,\min}<\omA<\omA^{\kapth,\max}$ and $\omA^{\kapmu,\min}<\omA<\omA^{\kapmu,\max}$, respectively. The maximum Alfv\'en frequencies for all modes are capped at $\omA^{\rm TI,\max}=\om$.

\begin{figure*}
    \centering
    \includegraphics[width=\linewidth]{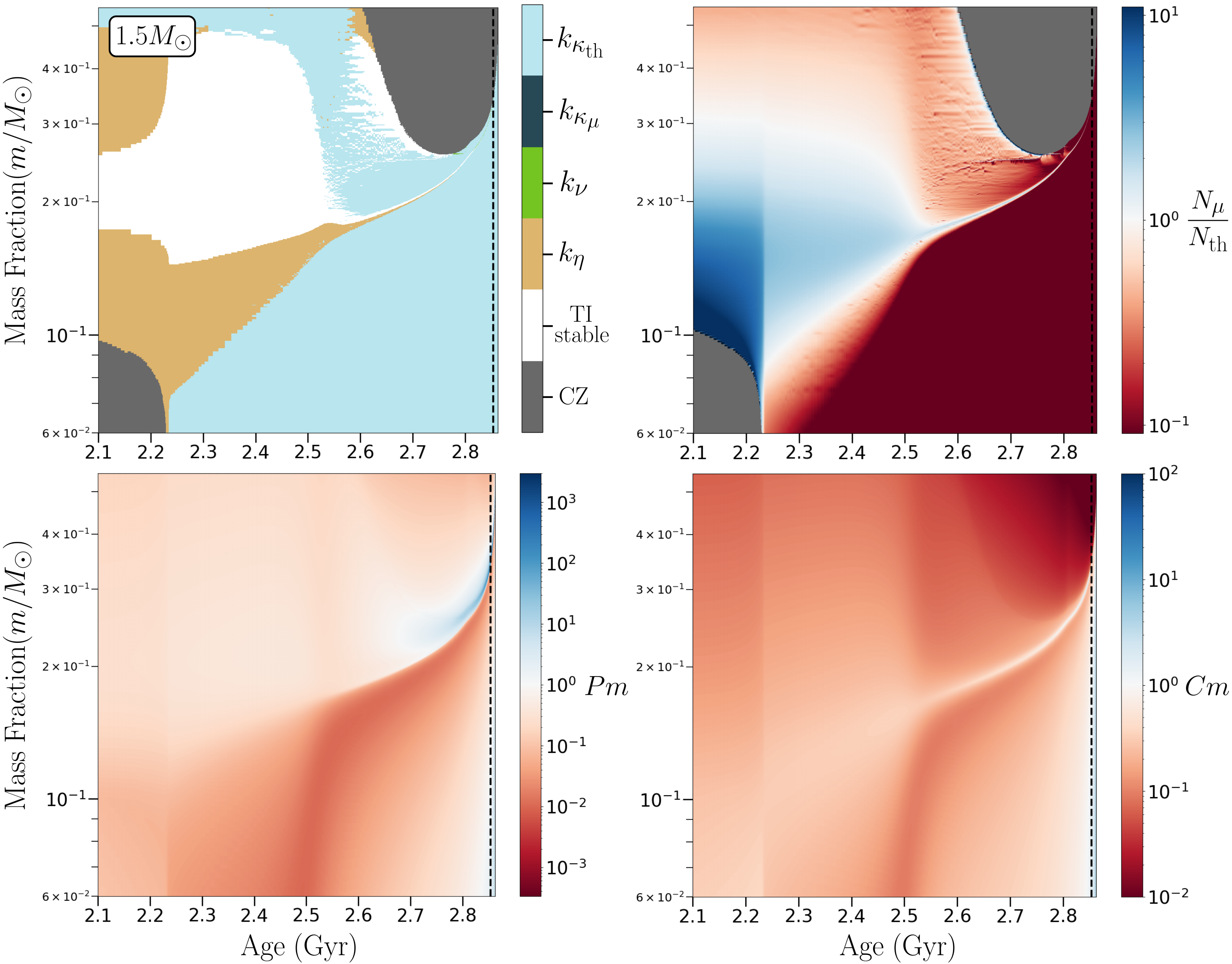}
    \caption{Map of TI modes and three key dimensionless parameters during the evolution of a $1.5M_\odot$ star. 
    Top left: the most unstable mode of the TI ($k_\nu$, $k_\eta$, $k_{\kapth}$, or $k_{\kapmu}$) is identified and indicated by color for each mass shell of the star excluding the convection zone (gray). The TI is suppressed in the white region.  
    Top right: $\Nmu/\Nth$, a proxy for the relative strength of compositional stratification. 
    Bottom left and right: $Pm=\nu/\eta$ and $Cm=\kapmu/\eta$. These parameters determine which TI modes can be unstable. Only the inner mass shells $m<0.55M_\odot$ are shown, from the end of the main sequence through the RGB phase. The edge of the growing helium core at $t\gtrsim 2.5$\,Gyr is approximately tracked by the thin strip where compositional stratification is dominant, $\Nmu/\Nth>1$. 
    }
    \label{fig:1p5M_proxies}
\end{figure*}

\begin{figure}
    \centering
    \includegraphics[width=\linewidth]{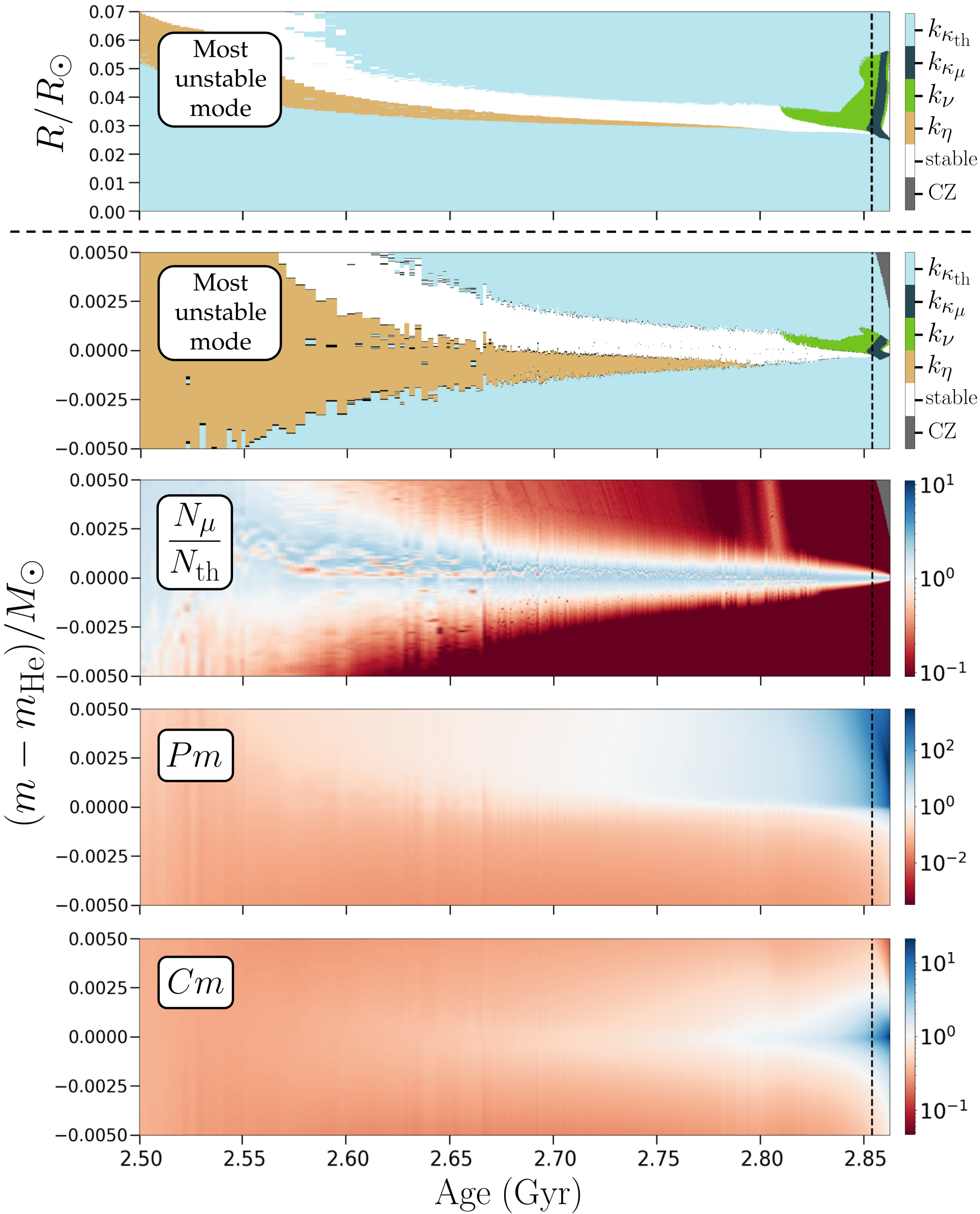}
    \caption{
    Zoom-in of the region around the helium core during the post-main sequence in the fiducial $1.5M_\odot$ model. Top two panels display the most unstable TI mode, shown on the $t$-$R$ and $t$-$m$ planes. In the white layer, the TI is disabled at all four canonical wavenumbers $k_\nu$,  $k_\eta$, $k_{\kapth}$, and $k_{\kapmu}$; this suppression occurs due to the strong compositional gradients $\Nmu/\Nth>1$ and $Cm<1$. Bottom three panels show $\Nmu/\Nth$, $Pm$, and $Cm$. The core boundary $m_{\rm He}(t)$ is defined at the peak of nuclear shell burning.
    }    \label{fig:1p5M_zoom_mHe}
\end{figure}
The values of $\omA^{\rm TI,\min}$  and $\omA^{\rm TI,\max}$ are determined by the set of parameters $\{\Nth,\Nmu,\Omega, R, k_\theta, \nu, \eta, \kapth, \kapmu\}$. The parameters relating to stellar structure $\{\Nth,\Nmu,\Omega, R\}$ are provided by MESA. The latitudinal wavenumber is a free parameter that we estimate with the strict lower bound $k_\theta=1/R$. This estimate will give the lowest values of the unstable intervals of $\omA$ for the $k_{\kapth}$ and $k_{\kapmu}$ modes (larger values of $k_\theta$ shift the instability intervals to higher $\omA$, since $\omA^{\rm TI,\min},\omA^{\rm TI,\max}\propto k_\theta^{1/2}$). For the microphysical diffusivities $\{\nu, \eta, \kapth, \kapmu\}$, we implement the standard expressions used in studies of convective zones \citep{jermyn2022atlas} and thermohaline mixing \citep{denissenkov2010numerical,wachlin2011thermohaline,garaud2015excitation}, as detailed in Appendix~\ref{ap:Diffusivities}. 

We define a mode to be unstable if
\begin{align}
    \label{eq:alpha}
    &\frac{\omA^{\rm TI,\max}}{\omA^{\rm TI,\min}}>\alpha,
\end{align}
where $\alpha\geq1$ is of order unity. A strict definition of the instability threshold would use $\alpha=1$, however in practice it is more convenient to use $\alpha$ somewhat larger than unity.\footnote{ For $\omA^{\rm TI,\min}=\omA^{\rm TI,\max}$, we have numerically found that the growth rate of each mode $\kTI$ is either shut off, $\gamma(\kTI)<0$,  or suppressed, $\gamma(\kTI)\ll \gamma^{\max}$, at all $\omA$. The growth rate attains its characteristic maximum value $\gamma^{\max}= \omA^2/4\Omega$ only when $\omA^{\rm TI,\min}\ll\omA\ll \omA^{\rm TI,\max}$, which requires $\alpha\gg 1$. In our analysis below, we find that increasing $\alpha$ primarily causes the regions of instability at $k_\eta$ and $k_{\kapmu}$ (in Figures~\ref{fig:1p5M_proxies} and \ref{fig:1p5M_zoom_mHe}) to recede since their unstable intervals of $\omA$ turn out to be the most narrow.}
We do not find qualitative differences in our results for $1\lesssim\alpha\lesssim 3$. Hereafter we use $\alpha=2$.

\subsection{Fiducial $1.5M_\odot$ model}
\label{fiducial}

Our fiducial stellar model has a mass of $1.5M_\odot$ and initial rotation with a surface speed of $50\;\rm{km}/\rm s$ at ZAMS. It is representative of a set of models with initial masses $(1-2)M_\odot$ and speeds $(25-200)\,\rm{km}/\rm s$ that have similar properties regarding the TI. We track the evolution of the star through the RGB phase until the helium flash in the core, which happens at $t=2.86$\,Gyr.

We begin by identifying which of the four TI modes ($k_\nu$, $k_\eta$, $k_{\kapth}$, or $k_{\kapmu}$) is most unstable in each mass shell of the star and at different phases of stellar evolution (Figure~\ref{fig:1p5M_proxies}). By definition, the most unstable TI mode means instability with the lowest threshold $\omA^{\rm TI,\min}$  that satisfies \Eq~(\ref{eq:alpha}). Figure~\ref{fig:1p5M_proxies} also shows the evolution of three key parameters $\Nmu/\Nth$, $Pm$, and $Cm$. They serve as useful proxies that help to quickly identify the regions of different TI modes and also conveniently track evolution of the core-envelope boundary. 

\begin{figure}
    \centering
    \includegraphics[width=\linewidth]{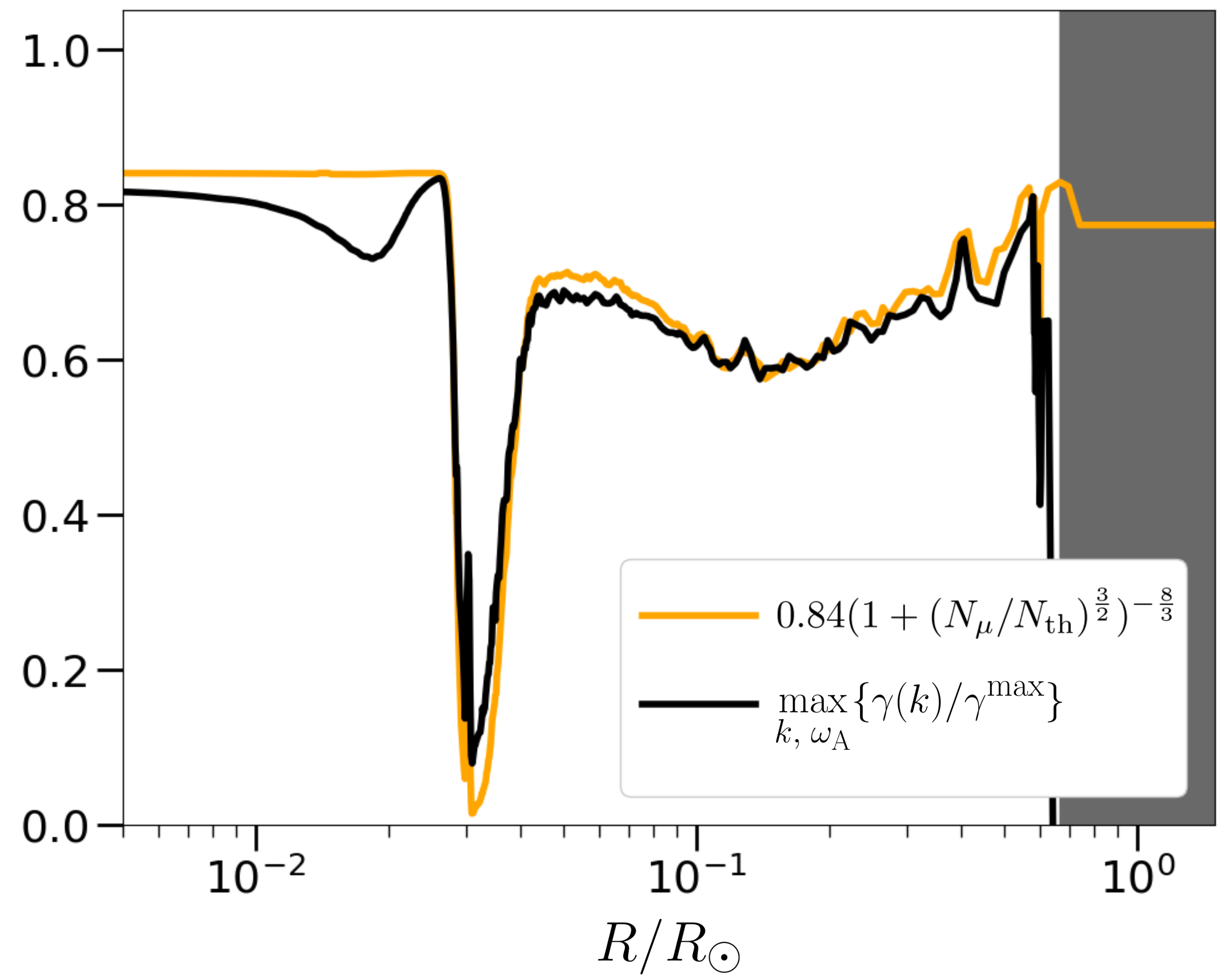}
    \caption{
    Radial profile of the maximum possible growth rate of the TI (normalized to the canonical $\gamma^{\max}=\omA^2/4\Omega$) in the $1.5M_\odot$ star with
    $m_{\rm He}= 0.25M_\odot$ (age $t=2.8\;$Gyr). The growth rate is obtained from numerical solutions of the dispersion relation at each radius $R$. It depends on $\omA$ as a parameter, and the black curve shows the maximum possible $\gamma$ found by scanning the interval of $0<\omA<\Omega$. 
    The deep pit observed outside the helium core $R_{\rm He}\approx 0.03R_\odot$ is in the region of strongest compositional stratification $\Nmu/\Nth\gtrsim 1$. The suppression of the growth rate is well reproduced by the simple expression $\gamma/\gamma^{\max}\approx 0.84/(1+(\Nmu/\Nth)^{3/2})^{-8/3}$, which matches onto the analytical estimate $\sim (\Nmu/\Nth)^{-4}$ derived in Appendix~\ref{ap:stablelayer} in the limit $\Nmu/\Nth\gg1$.
    Note that the maximum growth rate can only reach $\gamma\approx0.84\, \gamma^{\max}$ for the buoyancy-enabled TI modes $k_{\kapth}$ and $k_{\kapmu}$ (the prefactor $0.84$ is explained in SB24). Gray region at $R\gtrsim 0.7R_\odot$ indicates the convective envelope. 
    }
    \label{fig:gammamax_radialslice}
\end{figure}

First, consider the star near the end of the main sequence, $t\lesssim 2.2$\,Gyr. Its stably stratified zone consists of mass shells outside the convective core, $m\gtrsim 0.1 M_\odot$. Thermal stratification dominates in the outer shells: $\Nmu/\Nth<0.5$ at $m\gtrsim 0.5M_\odot$. Here, the MW instability at $k_{\kapth}$ (enabled by thermal diffusion) has the lowest threshold. Compositional gradients make a significant contribution to the stable stratification, $\Nmu/\Nth\gtrsim 0.5$, in the region $0.1\lesssim m/M_\odot\lesssim 0.5$. In this region, $Pm<1$ and $Cm<1$ (i.e. the magnetic diffusivity is dominant, $\eta>\nu,\kapmu$), so all MW modes are stable and only IW can be unstable, which corresponds to the $k_\eta$ mode of TI (enabled by magnetic diffusion). Therefore, the TI in the compositionally stratified region occurs if $\omA$ reaches the high threshold $\omAeta=\om\, Pm^{1/2}$. The instability at $k_\eta$ disappears in the middle of the compositionally stratified region 
(the white zone around $m=0.2M_\odot$ in Figure~\ref{fig:1p5M_proxies}) where the threshold becomes too high. 

Next, consider the post-main sequence phase and focus on the most interesting region: the core-envelope boundary where a strong compositional gradient is sustained across the hydrogen burning shell. At $t\gtrsim2.5$\,Gyr this region is narrow in the mass coordinate $m$; it is easily identified as the layer with $\Nmu/\Nth\gtrsim 1$ (see the top right panel of Figure~\ref{fig:1p5M_proxies}). The zoom-in of the boundary region is shown in Figure~\ref{fig:1p5M_zoom_mHe}, where one can see the key feature: a layer where all four canonical modes of TI are suppressed (the white strip). It persists at the core-envelope boundary in the evolving star, except for a brief period at the end of the RGB phase, around $t=2.85$\,Gyr. The suppression layer is narrow in the mass coordinate, $\Delta m/m_{\rm He}\sim 10^{-2}$, but has a significant width in radius, $\Delta R/R_{\rm He}\sim 0.2$, where $m_{\rm He}$ and $R_{\rm He}$ are the mass and radius of the helium core. The TI suppression can be traced to the values of $Pm$ and $Cm$. The modes at $k_\nu$ and $k_\eta$ are stable because neither $Pm\gg1$ nor $Pm\ll1$ is satisfied ($Pm\sim1$ in the layer), while $k_{\kapmu}$ is stable because $Cm\gg 1$ is not satisfied ($Cm\sim 1$ in the layer). The remaining canonical mode at $k_{\kapth}$ is necessarily stable since $\Nmu/\Nth>1$. 

The suppression of instability at all four canonical wavenumbers implies that the TI cannot develop with the usual growth rate $\gamma^{\max}=\omA^2/4\Omega$ in the compositionally stratified layer around the helium core. 
We now investigate the remaining possibility of a weak TI at non-canonical wavenumbers (with a growth rate $\gamma\ll \gamma^{\max}$) and include all six branches of the dispersion relation. In particular, we examine in detail the evolved star with a core mass $m_{\rm He}=0.25M_\odot$  (age $t=2.8$\,Gyr). At each radius, we scan the entire $k$ space by numerically solving the full dispersion relation (Appendix \ref{ap:LIA}) and checking its six roots $\omega(k)=\omega_r+i\gamma$ for $\gamma>0$. This brute-force approach identifies the fastest growing mode, if an instability exists. The result depends on $\omA$ as a parameter, and we have scanned the relevant interval $\omA<\Omega$ to find the maximum possible growth rate $\max_{k,\omA}\{\gamma(k)\}$. It is shown in Figure \ref{fig:gammamax_radialslice} as a function of radius $R$.

One can see that the maximum possible growth rate is suppressed in the compositionally stratified layer around the core boundary. It is reduced below $\gamma^{\max}$ by the factor $\sim(\Nmu/\Nth)^{-4}\ll 1$, as shown numerically in the figure and explained analytically in Appendix~\ref{ap:stablelayer}. The surviving weak instability occurs on the MW branch (the fifth and sixth branches are stable since $\kapmu\ll\eta$ is not satisfied, as $Cm\sim1$ in the layer). While the usual TI with $\gamma\approx \gamma^{\max}$ at the canonical wavenumber $k_{\kapth}$ is shut off, the remaining instability is found at $k_{\Nmu}$ with the maximum $\gamma\approx \gamma^{\max}(k_{\Nmu}/k_{\kapth})^{-4}$. The suppression factor $(k_{\Nmu}/k_{\kapth})^{-4}$ depends on $\omA$ and sharply peaks when $\omA=\omega_{\rm th}$, which gives the maximum $\gamma\sim \gamma^{\max} (\Nmu/\Nth)^{-4}$.

We note that although the suppression increases the TI growth timescale $\gamma^{-1}\gg (\gamma^{\max})^{-1}$,  it does not make $\gamma^{-1}$ exceed the evolutionary timescale $\sim10^8$\,yr. This may be seen using the estimate $(\gamma^{\max})^{-1}\sim 100$\,yr for typical parameters $\Omega=10^{-5}\;\rm s^{-1}$ and $\omA/\Omega=10^{-2}$. Thus, one can expect the TI to operate at some level in the layer with strong compositional stratification.

\begin{figure}
    \centering
    \includegraphics[width=\linewidth]{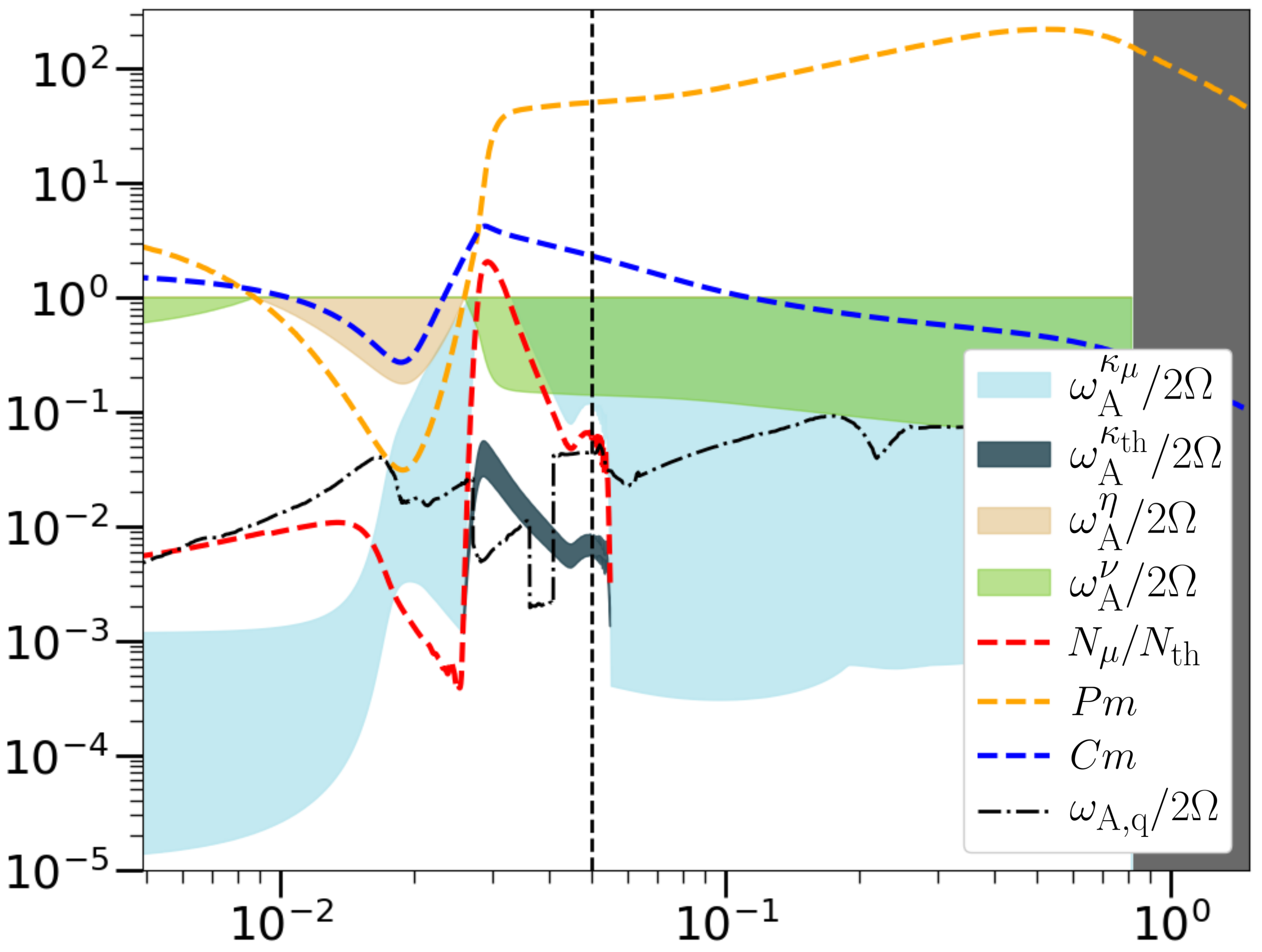}
    \includegraphics[width=\linewidth]{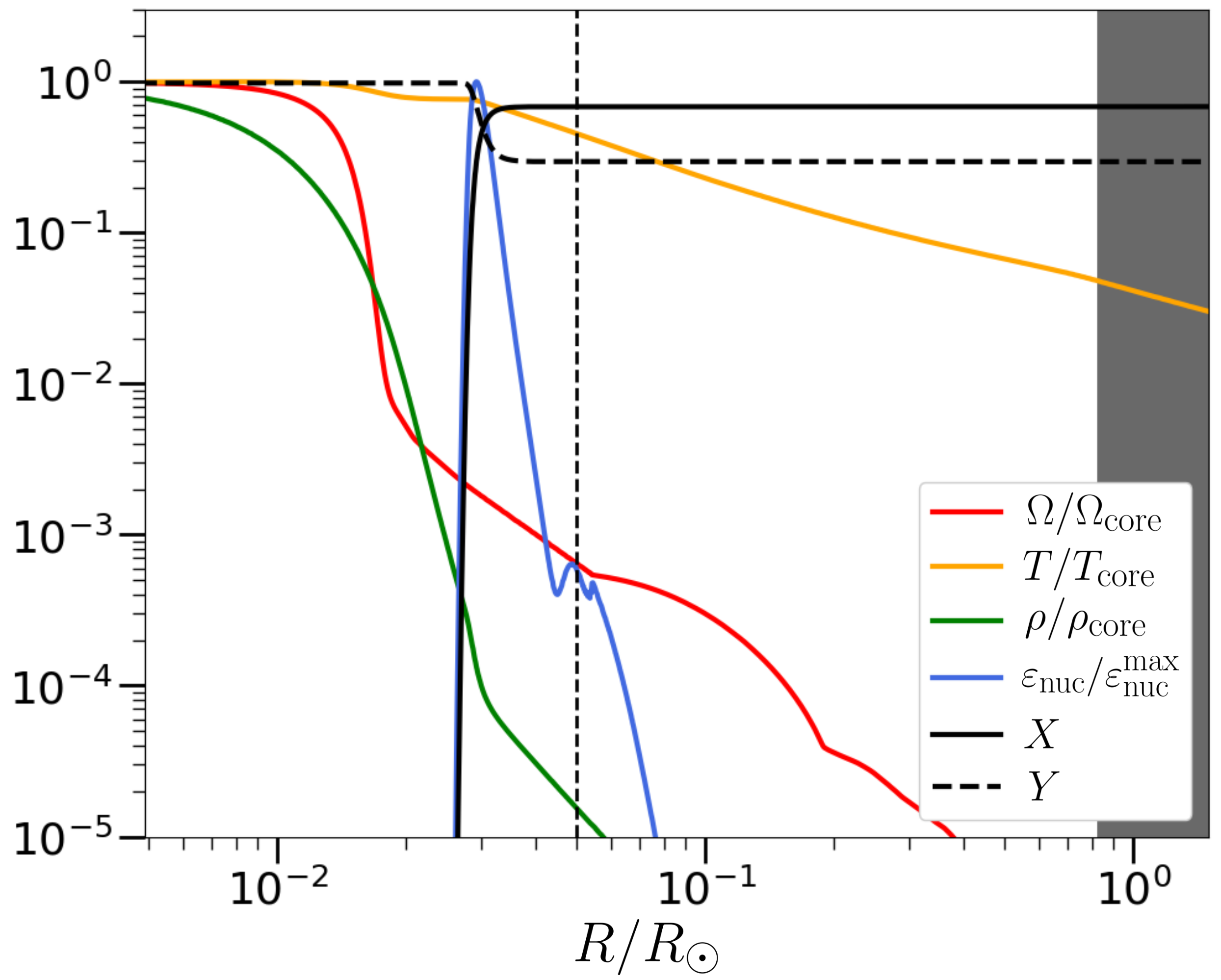}
    \caption{
    Radial profiles of a $1.5M_\odot$ star on the RGB when the helium core mass is $m_{\rm He}\approx 0.33M_\odot$. This stellar model is marked by the vertical dashed line at $t\approx 2.852$\,Gyr in Figures~\ref{fig:1p5M_proxies} and \ref{fig:1p5M_zoom_mHe}. Gray region at $R\sim R_\odot$ indicates the convective envelope. 
    Bottom: basic structure of the star, including the profiles of density $\rho$, temperature $T$, nuclear burning rate $\varepsilon_{\rm nuc}$, and the hydrogen and helium mass fractions $X$ and $Y$. 
    Top: instability intervals of $\omA/\om$ for each possible TI mode $\kTI=k_\nu,k_\eta,k_{\kapth}$, and $k_{\kapmu}$.
    Note that some of the instability intervals, in particular that of $k_{\kapmu}$, have width $\alpha<2$ at some radii and therefore do not appear in Figure~\ref{fig:1p5M_zoom_mHe} at this time (instability of $k_{\kapmu}$ becomes more robust at slightly later time $t\gtrsim 2.852$\,Gyr as $Cm$ increases). Dashed colored curves show the dimensionless parameters $\Nmu/\Nth$, $Pm$, and $Cm$, which control the TI. Black dash-dotted curve shows $\omega_{{\rm A},q}$ below which the TI analysis neglecting differential rotation is invalid. 
    The TI growth rate $\gamma(k)$ at a chosen radius $R=0.05R_\odot$ (vertical black dashed line) is shown for various $\omA$ in Figure~\ref{fig:1p5M_numericalsolver}.
    }
    \label{fig:1p5M_radialslice}
\end{figure}

When the helium core mass increases above $0.25M_\odot$ ($t\gtrsim2.8$\,Gyr), $Pm$ and $Cm$ both sharply rise in the burning shell to $Pm\sim 10^2$ and $Cm\sim 5$. Their values $Pm>Cm\propto T^4\rho^{-1}$ are controlled by the plasma temperature $T$ and density $\rho$ (one can get their dependence on $T$ and $\rho$ from the Spitzer scalings $\nu>\kapmu\propto T^{5/2}\rho^{-1}$ and $\eta\propto T^{-3/2}$ up to logarithmic corrections). As the helium core mass increases and its radius contracts, the density significantly drops at the core edge (connecting to the tenuous, extended envelope) while the temperature remains relatively constant. This causes the viscous and compositional diffusivities to increase relative to the magnetic diffusivity, so both $Pm$ and $Cm$ increase.

The increasing $Pm$ and $Cm$ at $t\gtrsim2.8$\,Gyr imply that the MW can become unstable at $k_\nu$ and $k_{\kapmu}$. This leads to a brief period in the star's life when canonical TI modes can operate throughout the stably stratified zone, including the compositionally stratified layer between the helium core and the hydrogen envelope. The onset of TI at wavenumbers $k_\nu$ and $k_{\kapmu}$ at $t\gtrsim 2.8$\,Gyr is seen in Figure~\ref{fig:1p5M_zoom_mHe} (green and dark blue shaded regions in the top two panels).

\begin{figure}
    \centering
    \includegraphics[width=\linewidth]{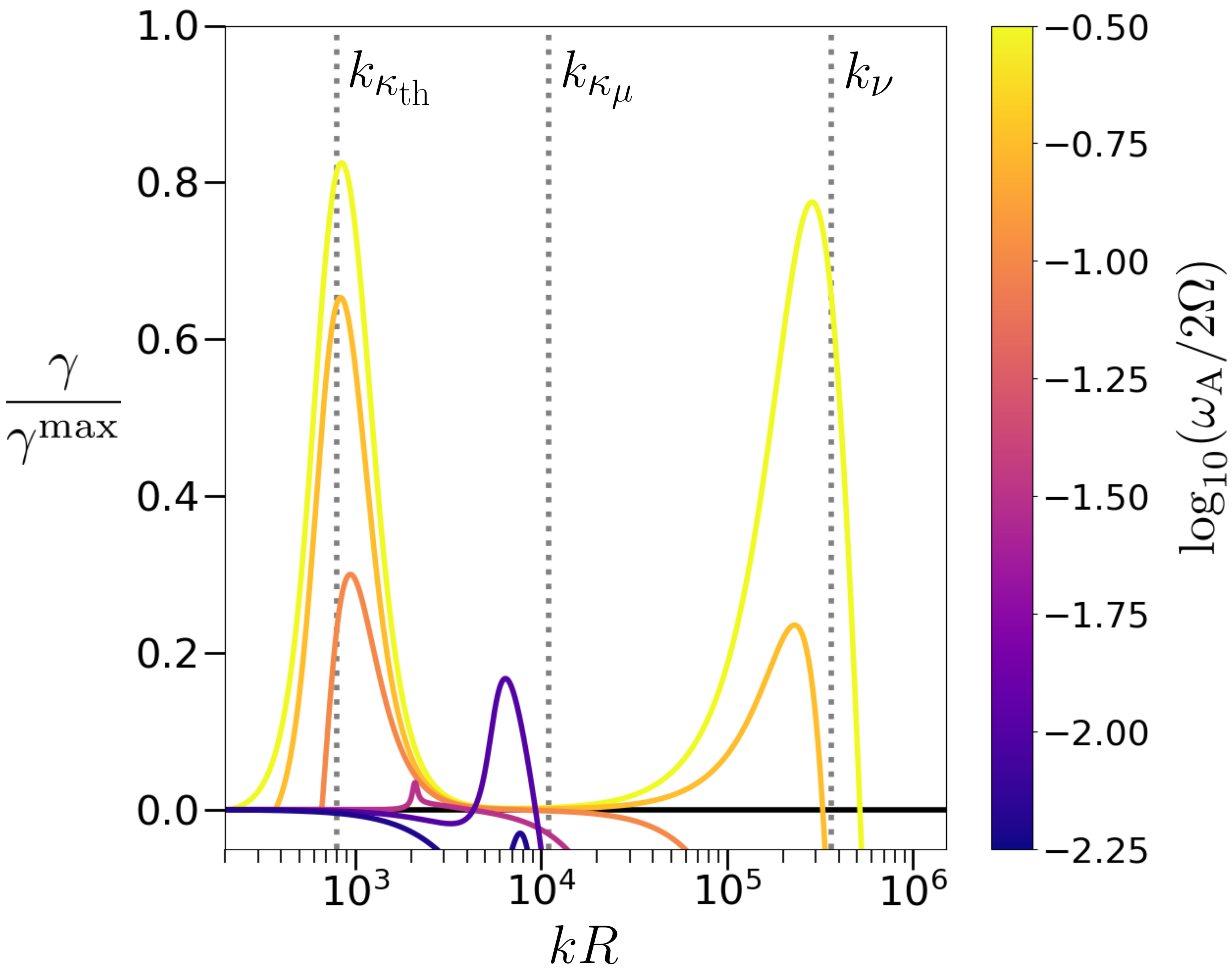}
    \caption{Numerical solution for the growth rate $\gamma(k)$ vs wavenumber $k$ at $R/R_\odot=0.05$ in a $1.5M_\odot$ star at $t=2.852$\,Gyr (when $m_{\rm He}\approx 0.33M_\odot$). This radius is marked by the vertical dashed black line in Figure \ref{fig:1p5M_radialslice} and chosen because there all three $\kTI$ of the MW branch can be unstable. The color-coded curves $\gamma(k)$ correspond to different $\omA/\om$. 
    }
    \label{fig:1p5M_numericalsolver}
\end{figure}

We conclude that the TI can robustly operate throughout the core-envelope transition for a relatively brief period of $\sim 10^7$\,yr near the end of the RGB phase. This period is also interesting from the TI physics point of view; therefore, we discuss it in some detail below. 

In particular, it is instructive to examine a radial slice of the core-envelope boundary at age $t=2.852$\,Gyr (when $m_{\rm He}\approx 0.33M_\odot$), which is marked by the vertical black dashed line in Figures~\ref{fig:1p5M_proxies} and \ref{fig:1p5M_zoom_mHe}. Figure~\ref{fig:1p5M_radialslice} shows the radial profiles of fluid parameters (which rapidly change near the helium core edge $R_{\rm He}\approx 0.03R_\odot$), and the intervals of $\omA/\om$ that give instability, for each possible TI mode. One can see that $Pm$ and $Cm$ increase with radius at $R>0.02R_\odot$ (due to the steep drop of density) and exceed unity, leading to the onset of TI at $k_\nu$ and $k_{\kapmu}$ in the layer at $R\approx 0.03R_\odot$, which has a strong compositional stratification, $\Nmu/\Nth\gtrsim 1$. The most unstable TI mode is $k_{\kapmu}$, as its instability appears at the lower $\omA/\om$. Outside the region of $k_{\kapmu}$ instability, thermal stratification strongly dominates and $k_{\kapth}$ is easiest to destabilize. 

This special, brief period in the star's life, with the diversity of unstable TI modes, can be further studied by numerically solving the dispersion relation (Appendix \ref{ap:LIA}) at a chosen radius. The numerical solution also provides an accuracy test of our analytical instability criteria. Figure~\ref{fig:1p5M_numericalsolver} shows the MW growth rate $\gamma_{\rm MW}(k)$ versus wavenumber $k$ at $R/R_\odot=0.05$ (marked by the vertical black dashed line in Figure~\ref{fig:1p5M_radialslice}) for various values of $\omA$ and fixed $k_\theta=1/R$. With increasing $\omA$, instability first appears at $\omA/\om\approx 10^{-2.2}$ near the wavenumber $k_{\kapmu}$. This instability is enabled by compositional diffusion, and no other TI modes operate at low $\omA$. The narrow interval of instability $\omA^{\kapmu,\max}/\omA^{\kapmu,\min}\sim 2$ leads to a maximum growth rate $\gamma\approx 0.05\omA^2/\Omega$, below the canonical $\gamma^{\max}=0.25\omA^2/\Omega$. When $\omA/\om$ exceeds $\sim 0.1$, the instability at $k_{\kapmu}$ is gone, and now the TI operates near $k_\nu$ and $k_{\kapth}$. The change in the TI properties with increasing $\omA$ closely matches the prediction of the analytical criteria in Figure~\ref{fig:1p5M_radialslice}. 

A layer with TI suppression is reinstated around the core after $t=2.86$\,Gyr, just before the helium flash. The growing compositional stratification and viscosity at the boundary of the contracting core leads to an increasing $k_{\kapmu}$ and decreasing $k_\nu$, which disable each other when $k_\nu$ decreases below $k_{\kapmu}$.

\subsection{Validity of Assumptions}

Our linear stability analysis made several assumptions, including the WKB approximation and the neglect of differential rotation and radial magnetic fields. Here, we briefly examine how justified these assumptions are for the $1.5M_\odot$ stellar model discussed above.

\paragraph{WKB approximation} 
The WKB approximation holds if the radial wavelengths of the TI modes are much shorter than the length scales on which the background quantities vary. This condition is most challenging to satisfy near the core edge where the stellar structure changes rapidly. Examining Figure~\ref{fig:1p5M_radialslice}, one can see that the fastest varying quantities, such as $\Nmu/\Nth$, vary over radial scales $\Delta R\lesssim 10^{-2}R_\odot$. In this region, we find for the relevant TI modes (see Figure~\ref{fig:1p5M_numericalsolver}): $k_{\kapth}\Delta R/2\pi\gtrsim 30$, $k_{\kapmu}\Delta R/2\pi\gtrsim 300$ and $k_\nu\Delta R/2\pi\gtrsim  10^4$. This separation of scales by more than an order of magnitude justifies the WKB approximation. 

\paragraph{Differential rotation} 
MWs are unaffected by differential rotation if the toroidal field is sufficiently strong, so that $\omA>\omega_{A,q}$ (Section \ref{sec:Randq}). Figure~\ref{fig:1p5M_radialslice} shows that this condition can be satisfied somewhere within the unstable interval of $\omA$ for all $\kTI$ modes outside the radius $R\gtrsim 2\times 10^{-2}R_\odot$. The condition is mildly violated for the $k_{\kapth}$ mode in the deeper core where differential rotation is strong. However, this may change with self-consistent inclusion of AM transport, which would reduce the differential rotation.

\paragraph{Radial magnetic field}  
The stabilizing effect of the radial magnetic field on the TI is negligible if $B_R$ is below the threshold in Equation~(\ref{eq:radial_field}). For typical values in the compositionally stratified region, this condition requires
\beq
B_R\ll3 
\left(\frac{10^3}{\kTI R}\right)\left(\frac{\omA/\om}{10^{-2}}\right)\left(\frac{\rho R^2}{ 10^{20}}\right)^{1/2}\left(\frac{\om}{ 10^{-5}}\right) \mathrm{G}.
\eeq
This is a strong constraint on $B_R$, which can easily be violated in stars. Values of $B_R\gtrsim 3\times 10^4$\,G in the hydrogen-burning shell were recently inferred from astroseismology of red giant cores \citep{li2022magnetic,deheuvels2023strong,li2023internal}. Such strong fields may be left over from the main sequence when a dynamo operated in the convective core \citep{fuller2015asteroseismology,cantiello2016asteroseismic,bugnet2021magnetic,becerra2022evolution}. Then, at later evolution phases, the TI can be suppressed out to the mass shell $m_{\rm conv}$ of the maximum extent of the earlier core convection. Note that $m_{\rm conv}$ increases with stellar mass $M$. For stars with $M\gtrsim 1.5M_\odot$, the hydrogen-burning shell lies within a previously convective region for a significant fraction of the lower RGB phase \citep{cantiello2016asteroseismic}. In lower mass stars, the radial field strength is uncertain because it is likely determined by fossil fields, whose properties remain poorly understood (for a review, see \cite{braithwaite2017magnetic}).

\subsection{Massive stars}

\begin{figure}
    \centering
    \includegraphics[width=\linewidth]{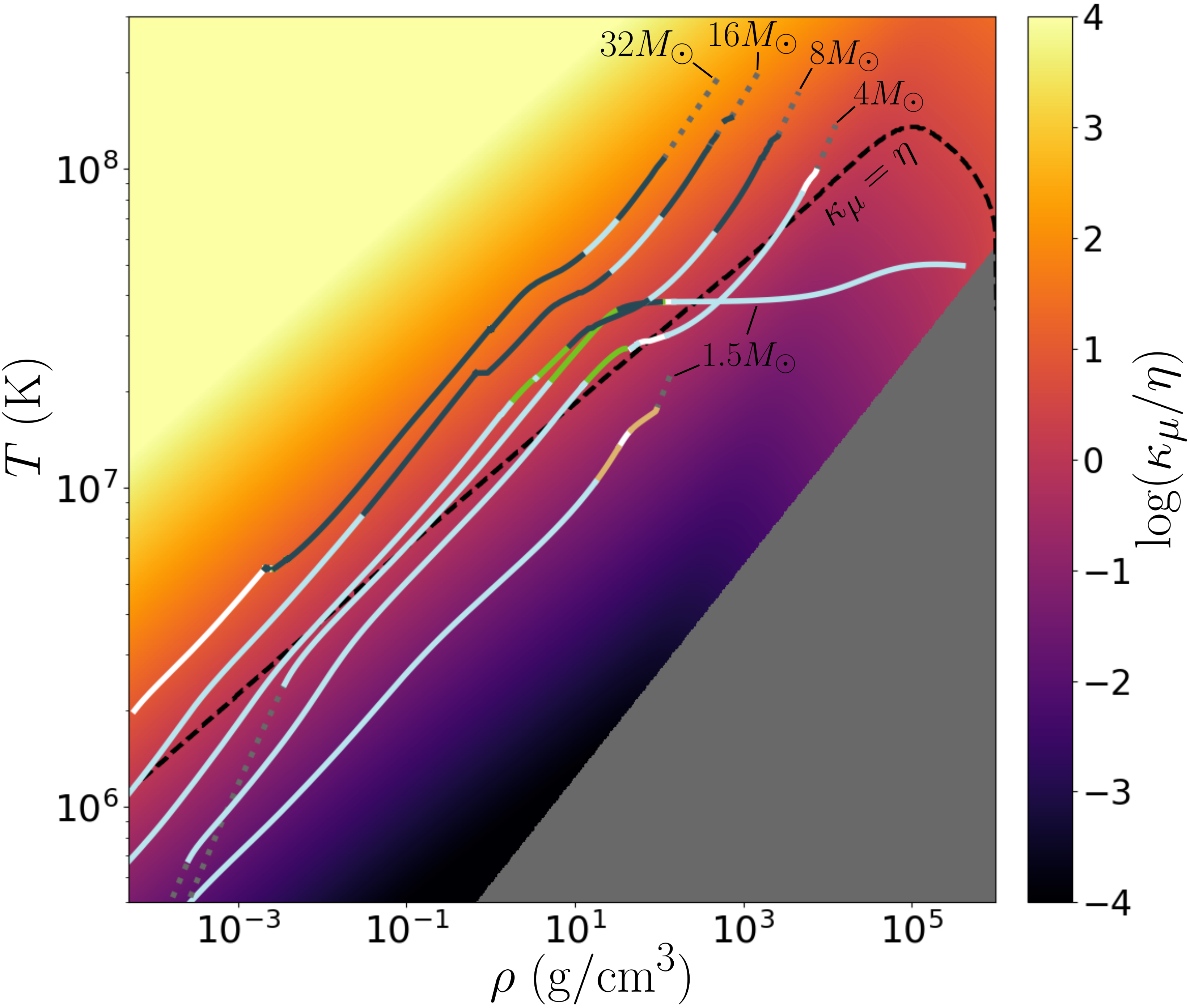}
    \caption{
    Stellar structure in the temperature-density plane against a color map of $Cm(T,\rho)=\kapmu/\eta$. Curves show the stellar structure of stars with $M/M_\odot=1.5,4,8,16$ and $32$. The $M=1.5M_\odot$ stellar model is shown at two evolution phases: during the main sequence ($t=2.1$\,Gyr, the lower curve) and the RGB phase ($t=2.85$\,Gyr). Stellar models with masses $4M_\odot$, $8M_\odot$, $16M_\odot$, and $32M_\odot$ are shown soon after the end of their main sequence, at ages $t=170$\,Myr, $34$\,Myr, $11$\,Myr and $5.4$\,Myr, respectively. All the curves are colored according to the most unstable TI mode (the color code is taken from Figure~\ref{fig:1p5M_proxies}), and their dotted gray portions represent convective regions. Stars more massive than $\sim4M_\odot$ have $Cm>1$ throughout their deeper interiors and the TI is typically not suppressed in compositionally stratified regions during their evolution. The $Cm$ color map was calculated for a H-He plasma with hydrogen mass fraction $X=0.7$. The region shaded in gray is where the Coulomb coupling parameter $\Gamma$ \citep{hubbard1966studies} is larger than unity and our estimates for $Cm$ are not applicable.}
    \label{fig:Cm}
\end{figure}

When analyzing the TI in more massive stars, the main change can be easily understood by examining $Pm=\nu/\eta$ and $Cm=\kapmu/\eta$, which control the TI modes as discussed above. The scaling $Pm>Cm\propto T^4/\rho$ implies that interiors of massive stars have $Pm>1$ and $Cm>1$ because of their high temperatures. In addition, at high temperatures, radiation makes a significant contribution $\nu_{\rm rad}$ to the viscosity, on top of the usual plasma viscosity due to ion transport, $\nu_{\rm ion}\sim \kapmu$. Therefore, the viscosity is increased to $\nu\sim \nu_{\rm rad}+\kappa_\mu$, increasing $Pm$ further. The condition $Pm\gg 1$ implies the stability of the IW branch and leaves three possible TI modes of the MW branch: $k_{\kapth}$, $k_{\kapmu}$, and/or $k_\nu$.

The trend of increasing $Cm$ in massive stars is demonstrated in Figure~\ref{fig:Cm}, which compares stars with $M/M_\odot=1.5,4,8,16,$ and $32$. One can see that the lower mass stars, $M=1.5M_\odot$ and $4M_\odot$, have $Cm<1$ everywhere except in layers at the edge of the helium core that develops during post-main sequence evolution. Stars with mass $M\gtrsim 4M_\odot$ have $Cm>1$ throughout their deep interiors. Therefore, the TI can develop everywhere in a massive star, including its deep layers with compositional stratification. Typically, either $k_{\kapth}$ or $k_{\kapmu}$ is the most unstable mode, depending on whether the local stratification is predominantly thermal or compositional. In some cases, a large $Pm\gg Cm$ causes $k_\nu$ to be the easiest to destabilize when $\omAnu\propto Pm^{-1/2}$ becomes smaller than $\omA^{\kapth,\min}$ and $\omA^{\kapmu,\min}$. 

\section{Conclusions}
\label{sec:Discussion}

Our results have extended the linear analysis of the TI to cover all regimes encountered in stellar interiors. In particular, we have generalized the analysis of \SB~to include stratification with both compositional and thermal components.  We find that each of the diffusivities in a stellar plasma (viscous $\nu$, magnetic $\eta$, thermal $\kapth$, and compositional $\kapmu$) can enable the TI with a maximum growth rate near $\gamma^{\max}= \omA^2/4\Omega$ at four canonical wavenumbers $\kTI$ where the associated diffusive timescale matches the rotational timescale $\Omega^{-1}$. Viscous, thermal, and compositional diffusion destabilizes the MW branch at wavenumbers $k_\nu$, $k_{\kapth}$, and $k_{\kapmu}$, respectively, while magnetic diffusion destabilizes the IW branch at $k_\eta$. We formulated analytical stability criteria for all modes, allowing for a straightforward implementation of a ``toggle switch" for the TI with the canonical growth rate $\sim \gamma^{\max}$ that is assumed in existing models of AM transport due to the Tayler-Spruit dynamo.

We have implemented such a toggle switch in the MESA stellar evolution code to broadly examine the stability of TI modes in stellar interiors. We find that low and high mass stars have qualitatively different stability patterns due to different internal profiles of the diffusivities. In thermally stratified regions, at least one of the TI modes (enabled by either viscous, magnetic or thermal diffusion) can be active in all stars. However, in regions with strong compositional stratification the TI relies on the slow compositional diffusion and requires a relatively demanding condition $Cm= \kapmu/\eta>1$. 

Since $Cm$ steeply increases with temperature ($Cm\propto T^4$), the TI easily develops throughout the deep interiors of hot, high-mass stars. For stars of lower mass, which have lower internal temperatures, we find that the TI is suppressed in part of the compositionally stratified layers.\footnote{We note that more extreme toroidal field configurations with large gradients $p\equiv \partial \ln B_\phi/\partial \ln r>3/2$ can destabilize the MW branch even in the compositionally stratified regions with $Cm<1$ (Appendix \ref{ap:Large_p}). However, it is unclear how such a configuration would form in a star; therefore, this and previous works (e.g. \cite{spruit1999differential,zahn2007magnetic,ma2019angular}) focus on configurations with moderate gradients $p<3/2$.} These layers are of significant interest because they are located in the transition region of differential rotation between the core and the envelope of evolved stars. We find that in this zone the TI growth rate is suppressed by at least a factor of $(\Nmu/\Nth)^4$ below $\gamma^{\max}= \omA^2/4\Omega$, and that this suppression persists through most of the RGB phase. We leave a detailed examination of the impact of suppressed TI growth rates on the level of turbulent transport in these layers for future work.

In addition, we note that the expected strong poloidal magnetic fields left over in stars with main-sequence convective zones \citep{cantiello2016asteroseismic} can easily prevent the TI. Recent observations inferring  $B_R\gtrsim 3\times 10^4$\,G fields \citep{li2022magnetic,deheuvels2023strong,li2023internal} are well above the threshold $B_R\sim 3$\,G needed to suppress the TI in the compositional layer of a $1.5M_\odot$ star. The stabilizing role of remnant poloidal fields is an issue for a self-consistent modeling of AM transport with the Tayler-Spruit dynamo. In an accompanying paper, we argue that stable magnetic configurations are ubiquitous in stars due to their memory of extinguished convective zones, and they form ``magnetic webs'' that resist differential rotation and greatly assist AM transport in radiative zones \citep{skoutnev2025magnetic}.

\begin{acknowledgments}
We thank Matteo Cantiello, Jared Goldberg, Jim Fuller, Brian Metzger, and Mathieu Renzo for helpful discussions. This work is supported by NSF grant AST-2408199. A.M.B. is also supported by NASA grant 21-ATP21-0056 and Simons Foundation award No. 446228.
\end{acknowledgments}

%

\vspace{5mm}





\appendix

\section{Linear stability analysis with multiple types of stratification}
\label{ap:LIA}

The linear stability of a toroidal magnetic field configuration in a rotating star with a single type of stable stratification was examined in \SB. In this Section, we briefly review the approximations made to derive the dispersion relation for wave-like perturbations and then generalize the result to two types of stable stratification.

The magnetized fluid in a stellar interior is described by the  MHD equations, which govern the velocity field, magnetic field $\boldsymbol{B}$, and thermodynamic variables. The focus here is on the stability of perturbations on top of a background toroidal field $B_\phi$ in a uniformly rotating, stably stratified star. The most unstable perturbations are nearly incompressible and have small length and velocity scales compared to the local scale height and sound speed, respectively \citep{tayler1973adiabatic,spruit1999differential}. In this limit, the MHD equations are simplified in the Boussinesq approximation \citep{spiegel1960boussinesq}, where, for a thermally and compositionally stratified fluid, the buoyancy force in a gravitational field is captured by linear contributions of the temperature and composition perturbations to the density perturbations, known as buoyancy variables. Pressure contributions to the density perturbations are second order and neglected.

In the WKB approximation, the dependence of perturbations on the spherical radius $R$ and polar angle $\theta$ can be approximated as local plane waves in the poloidal plane. Perturbations then take the form $\propto\exp[i(k_R R+l\theta +m\phi-\omega t)]$, where the frequency is complex $\omega=\omega_r+i\gamma$ and the wavevector is
\beq
  \bk=(k_R,k_\theta,k_\phi), \qquad k_\theta=\frac{l}{R}, \quad k_\phi=\frac{m}{r}.
\eeq
Unstable modes of the TI have short radial and long horizontal wavelengths to minimize the potential energy cost of radial motions against the background stratification \citep{spruit1999differential}. Wavevectors with short radial wavelengths satisfy
\beq
  k\approx k_R\gg k_\theta\gtrsim  k_\phi\sim\frac{1}{r}.
\eeq

Linear stability analysis proceeds by obtaining the dispersion relation $D(\omega,\boldsymbol{k})=0$ from the linear dynamical equations for perturbations. Its derivation with a single type of stratification (with Brunt-V\"ais\"al\"a frequency $N$ and buoyancy diffusivity $\kappa$) has been detailed in previous works; the result is
\begin{align}
   \label{eq:DispersionSpherical_singlestrat}
   D(\omega) =& \left(\omega_\nu\omega_\eta-m_\star^2\omA^2-\frac{k_\theta^2 N^2}{k^2}\frac{\omega_\eta}{\omega_\kappa}\right) \left(\omega_\nu\omega_\eta-m^2\omA^2\right)- 4\cos^2\! \theta \left(\Omega \omega_\eta+m\omA^2\right)^2=0,\\
    \label{eq:mstar}
    m_\star^2 \equiv&  m^2 - 2\cos\theta\, \frac{r}{R}\,\partial_\theta \ln \frac{B_\phi}{r} = m^2 - 2 (p\cos^2\theta - 1),\qquad \omega_s\equiv \omega+isk^2, 
    \qquad s\in\{\nu,\eta,\kappa\},
\end{align}
where $N=\Nth$ or $N=\Nmu$, and $\kappa=\kapth$ or $\kappa=\kapmu$, depending on the type of stratification.

When both types of stratification are present, the MHD perturbation equations change in a straightforward way. There are now two equations describing the two buoyancy variables, thermal and compositional. Since the net buoyancy force is the sum of the thermal and compositional contributions, the stratification term in the dispersion relation simply becomes a sum of the corresponding terms proportional to $\Nth^2$ and $\Nmu^2$, so \Eq~(\ref{eq:DispersionSpherical_singlestrat}) changes to
\begin{align}
   \label{eq:DispersionSpherical_doublestrat}
   D(\omega) =& \left(\omega_\nu\omega_\eta-m_\star^2\omA^2-\frac{k_\theta^2 \Nth^2}{k^2}\frac{\omega_\eta}{\omega_{\kapth}}-\frac{k_\theta^2 \Nmu^2}{k^2}\frac{\omega_\eta}{\omega_{\kappa_{\mu}}}\right) \left(\omega_\nu\omega_\eta-m^2\omA^2\right)- 4\cos^2\! \theta \left(\Omega \omega_\eta+m\omA^2\right)^2=0.
\end{align}

To simplify the analysis and focus on the effects of two types of stratification, we will consider only polar regions $|\cos\theta|\approx 1$ where modes are most unstable. By Stokes theorem, the field there must have a latitudinal dependence $p=\partial\ln B_\phi/\partial\ln r=1$ for any finite current along the rotation axis. Then, only perturbations with azimuthal modes $m=1$ are unstable (\SB), and the dispersion relation is simplified to

\begin{align}
\label{eq:DispersionSimple_doublestrat}
    D(\omega)  &=\left(\omega_\nu\omega_\eta-\omA^2-\frac{k_\theta^2 \Nth^2}{k^2}\frac{\omega_\eta}{\omega_{\kapth}}-\frac{k_\theta^2 \Nmu^2}{k^2}\frac{\omega_\eta}{\omega_{\kappa_{\mu}}}\right)
    \left(\omega_\nu\omega_\eta-\omA^2\right)
    -\left(2\Omega \omega_\eta+2\omA^2\right)^2=0.
\end{align}

In the case of a single stratification type, $D(\omega)$ is a fifth order polynomial (with complex coefficients). In the case of two types of stratification, $D(\omega)$ is a sixth order polynomial, with six complex roots. Solutions for $\omega$ with a positive imaginary component $\gamma(k)>0$ are unstable. The solutions can be classified as follows. The small parameter $\omA/\om\ll1$ in a rotating star leads to two types of waves: IW with high frequencies $|\omega|\sim \om$ and MW with low frequencies $|\omega|\sim\omA^2/\om$. There are two IW and two MW solutions, accounting for four of the six roots. 

The remaining fifth and sixth roots have nearly zero real frequency $\omega_r\approx 0$ in the ideal MHD limit (also called direct modes in \cite{zahn2007magnetic}). \SB~analyzed stability of the fifth root $\omega_5$ in the case of a single type of stratification and found that the growth rate $\gamma_5$ is much smaller than $\omA^2/4\Omega$ when $\omA$ is larger than its instability threshold. The growth rate only reaches $\lesssim\omA^2/4\Omega$ in a narrow interval of $\omA^2/2\Omega\sim \eta k_N^2$. Similar properties are shared by the fifth and sixth roots in the case of two types of stratification. Note that instabilities of these modes require $\kapth$ or $\kapmu$ to be smaller than the magnetic diffusivity, which can only occur for $\kapmu$ in stars. Since the growth rates of the fifth and six roots do not grow with the canonical growth rate $\omA^2/4\Omega$ used in models of the Tayler-Spruit dynamo, we only consider them in this paper when all four of the canonical modes are stable.

\section{Instability of Inertial Waves}
\label{ap:IWs}
Rotating stars with $\Omega\gg\omA$ contain IW modes with high oscillation frequencies $\omega_r\sim2\Omega\gg\omA$. Magnetic diffusion can destabilize these modes. In \SB, we found that the IW growth rate $\gamma_{\rm IW}(k)$ reaches a maximum $\gamma_{\rm IW}^{\max}\approx \omA^2/4\Omega$ at the wavenumber $k_\eta=(\om/\eta)^{1/2}$ where magnetic diffusion and rotational timescales are comparable, $t_\eta\approx t_\Omega$. The expression for the maximum growth rate that includes the effects of viscosity and a single type of stratification can be obtained from \Eqs~(47) and (51) of \SB. The generalization to multiple types of stratification is straightforward:
\begin{align}\label{eq:gammaIW_resist}
    \gamma_{\rm IW}^{\max}=\frac{\omA^2}{4\Omega}-\sum_i\frac{\kappa_i k_\theta^2N_i^2}{8\Omega^2(1+\kappa_i^2/\eta^2)} -\sum_i\frac{\omA^2\eta k_\theta^2 N_i^2}{32\Omega^4(1+\kappa_i^2/\eta^2)}-\frac{2\Omega \nu}{\eta},
\end{align}
where the summation is over the two types of stratification: $i={\rm th},\mu$.
Instability $\gamma_{\rm IW}^{\max}>0$ is possible only if all three negative terms are small compared to the first term. Requiring the second term to be smaller than the first gives
\begin{align}
    \label{eq:IW_instcrit_1}
    \frac{\omA^2}{4\Omega^2}\gg \sum_i \frac{N_i^2}{4\Omega^2}\left(\frac{\kappa_i k_\theta^2}{2\Omega}\right)\left(1+\frac{\kappa_i^2}{\eta^2}\right)^{-1}.
\end{align}
Requiring the third term to be small compared to the first gives a condition on the stratification: 
\begin{align}\label{eq:IW_instcrit_2}
    &\sum_i\frac{N_i^2}{4\Omega^2}\left(1+\frac{\kappa_i^2}{\eta^2}\right)^{-1}\ll\left(\frac{\eta k_\theta^2}{\om}\right)^{-1} .
\end{align}
The last term being small compared to the first term can be written as another condition on $\omA$:
\begin{align}\label{eq:IW_instcrit_3}
    &\frac{\omega_A}{\om}\gg Pm^{1/2}.
\end{align}
Its physical interpretation is that the viscous diffusion rate needs to be slower than the instability growth rate $\nu k_\eta^2\ll \omA^2/4\Omega$. One can see that a small magnetic Prandtl number $Pm\ll1$ is required to support instability of IW.

In summary, the necessary and sufficient conditions for IW instability with maximum growth rate $\gamma_{\rm IW}^{\max}\approx \omA^2/4\Omega$ are given by the condition on the stratification in \Eq~(\ref{eq:IW_instcrit_2}) and the condition $\omA>\omega_{\min}$, where $\omega_{\min}$ is the largest of the lower limits in \Eqs~(\ref{eq:IW_instcrit_1}) and~(\ref{eq:IW_instcrit_3}). These results are summarized in the first row of Table \ref{tab:InstabilityCriteria_etaandnu}.

\section{Instability of Magnetostrophic Waves}
\label{ap:MWs}
Rotating stars with $\Omega\gg \omA$ support MW with low oscillation frequencies $\omega_r\sim\omA^2/\om\ll\omA$. Instability of MW can be independently enabled by both viscosity and the diffusive buoyancy response (either due to thermal or compositional stratification). Below, we first examine instability due to viscosity, then due to buoyancy effects. 

\subsection{MW instability enabled by viscosity }
\label{ap:MW_at_knu}
The growth rate of MW reaches a maximum $\gamma_{\rm MW}^{\max}\approx \omA^2/4\Omega$ at the wavenumber $k_\nu=(\om/\nu)^{1/2}$ where the viscous diffusion and rotational timescales are comparable $t_\nu\approx t_\Omega$. The maximum growth rate may be suppressed if buoyancy effects or magnetic diffusion are significant at $k_\nu$.

Let us first consider the effects of magnetic diffusion. The maximum growth rate at $k_\nu$ with neglected buoyancy effects ($N_i=0$) and small $\eta\ll\nu$ can be obtained from 
\Eq~(49) in \SB:
\begin{align}\label{eq:gammaMW_visc}
    \gamma_{\rm MW}^{\max}= \frac{\omA^2}{4\Omega}-\frac{2\Omega\eta}{\nu}.
\end{align}
Requiring magnetic diffusion (the second term) to be negligible compared to the maximum growth rate (equivalent to $\eta k_\nu^2\ll\omA^2/4\Omega$) can be written as a condition on $\omA$:
\begin{align}\label{eq:CriteriaInstMW_knu_1}
    \frac{\omA}{\om}\gg Pm^{-1/2}.
\end{align}
One can see that a large magnetic Prandtl number $Pm\gg 1$ is required for MW instability at $k_\nu$.

Next, consider the effects of buoyancy. Buoyancy affects modes with low wavenumbers and can reduce or suppress the maximum growth rate at $k_\nu$. The expression for the growth rate $\gamma_{\rm MW}(k)$ for $k<k_\nu$ that includes the effect of a single type of stratification is given by Equation~(60) in \SB. When evaluated at $k\sim k_\nu$, it approximately shows how the peak growth rate $\gamma_{\rm MW}(k_\nu)$ is affected by stratification. Its extension to multiple kinds of stratification is
\begin{align}\label{eq:lowMWnu_instcond_rewrite1}
    \displaystyle{\gamma_{\rm MW}(k_\nu)\approx\frac{\omA^2}{\om}\left[1-\sum_i\left(\frac{k_{\kappa_i}}{k_\nu}\right)^4\frac{\kappa_i}{\nu}\left(\frac{\omA^2}{4\Omega^2}-\frac{\kappa_i}{\nu}\right)\left(\frac{\omA^4}{16\Omega^4}+\frac{\kappa_i^2}{\nu^2}\right)^{-1}\right]>0,\quad k_{\kappa_i}\equiv \left(\frac{k_\theta^2 N_i^2}{\om\kappa_i}\right)^{1/4}.}
\end{align}
One can see that the growth rate at $k_\nu$ is unaffected by a component of the stratification ($N_i=\Nth$ or $\Nmu$) if $k_{\kappa_i}/k_\nu\ll1$. If $k_{\kappa_i}/k_\nu\gg1$, then the effect of stratification $i$ on $\gamma_{\rm MW}(k_\nu)$ can still be small, as long as
\begin{align}\label{eq:lowMWnu_instcond_comp}
    \displaystyle{\frac{\omA}{2\Omega}\gg\left(\frac{k_{\kappa_i}}{k_\nu}\right)^2\left(\frac{\kappa_i}{\nu}\right)^{1/2}=\frac{N_i}{2\Omega}\left(\frac{\nu k_\theta^2}{\om}\right)^{1/2},\quad \frac{k_{\kappa_i}}{k_\nu}\gg1,}
\end{align}
which is equivalent to $k_{N_i}\ll k_\nu$, where $k_{N_i}=k_\theta N_i/\omA$ (see discussion in \SB). Combining the results for both limits of $k_{\kappa_i}/k_\nu$, the condition for TI with $\gamma_{\rm MW}^{\max}(k_\nu)\approx \omA^2/4\Omega$ to be unaffected by stratification is
\begin{align}\label{eq:CriteriaInstMW_knu_2}
    \frac{\omA}{2\Omega}\gg\sum_i\Theta\left(\frac{k_{\kappa_i}}{k_\nu}-1\right)\frac{N_i}{2\Omega}\left(\frac{\nu k_\theta^2}{\om}\right)^{1/2},
\end{align}
where $\Theta(x)$ is the Heaviside step function.

In summary, the necessary and sufficient conditions for MW instability at wavenumber $k_\nu$ with the growth rate $\gamma_{\rm MW}^{\max}\approx \omA^2/4\Omega$ are given in \Eqs~(\ref{eq:CriteriaInstMW_knu_1}) and (\ref{eq:CriteriaInstMW_knu_2}), which are also stated the second row of Table~\ref{tab:InstabilityCriteria_etaandnu}.

\subsection{MW instability enabled by diffusive buoyancy}
\label{ap:MW_buoyancy}

As shown in SB24 for a single type of stratification, diffusive buoyancy effects in a stratified fluid can destabilize MW at wavenumbers $k_\kappa$ where the timescale for the diffusive buoyancy response is comparable to the rotation timescale, $t_\kappa\sim t_\Omega$. The instability behavior in the presence of multiple types of stratification is more complicated, and requires one to redo the analysis of the MW dispersion relation.

The MW roots of the dispersion relation (\Eq~\ref{eq:DispersionSimple_doublestrat}) satisfy $|\omega|\sim \omA^2/\om\ll\omA$. Thus, the higher order terms proportional to $\omega^4$, $\omega^3$, and $\omega^2\omA^2$ (related to the inertial response) can be dropped because they are small compared to the Alfve\'nic terms $\propto \omA^4$. We further neglect viscous diffusion (because it affects only high wavenumbers near $k_\nu$, as discussed in Section~\ref{ap:MW_at_knu}) but will include the effects of magnetic diffusion, treating them as a small correction. The dispersion relation in the magnetostrophic limit becomes
\begin{align}\label{eq:MW_dispersion_normalized}
    D(\omega)  &\displaystyle{=\hat\omega_\eta^2+4\hat\omega_\eta+3-\sum_i\frac{k_\theta^2 N_{i}^2}{k^2\omA^2}\frac{\hat\omega_\eta}{\hat\omega_{\kappa_i}}=0},
    \qquad
    \hat\omega_s\equiv \frac{\om}{\omA^2} \omega_s \quad (s\in\{\nu,\eta,\kapth,\kapmu\}). 
\end{align}
Note that the stratification terms $\propto N_i^2$ simplify in the limits of $\kappa_i k^2\ll |\omega|$ and $\kappa_i k^2\gg |\omega|$ (slow and fast buoyancy diffusion compared to the frequency of the mode $k$):
\begin{equation}\label{eq:strattermlimits}
 \displaystyle{\frac{k_\theta^2 N_i^2}{\omA^2 k^2}\frac{\omega_\eta}{\omega_{\kappa_i}}=\frac{\hat\omega_\eta}{\frac{k^2}{k_{N_i}^2}\hat\omega+i\frac{k^4}{k_{\kappa_i}^4}}\approx } \left\{\begin{array}{ccl}
    \displaystyle{\frac{k_{N_i}^2}{k^2}}, & \displaystyle{\quad \kappa_i k^2\ll
    |\omega|,}
    & \quad \displaystyle{k_{N_i}\equiv  k_\theta\frac{N_i}{\omA}} \\
   \displaystyle{-i\frac{k_{\kappa_i}^4}{k^4}\hat\omega_\eta,} & \displaystyle{\quad \kappa_i k^2\gg |\omega|,}
   & \quad \displaystyle{k_{\kappa_i}\equiv  \left(\frac{k_\theta^2 N_i^2}{\om\kappa_i}\right)^{1/4}}\\
             \end{array}\right.
\end{equation}
In the top row, we have used the assumption of weak magnetic diffusion to approximate $\omega_\eta/\omega\approx 1$. 

Before moving on to analyze solutions of \Eq~(\ref{eq:MW_dispersion_normalized}), it is helpful to briefly review its simpler version in the case of a single type of stratification,
\begin{align}\label{eq:MW_dispersion3}
  D(\omega)  &=\hat\omega_\eta^2+4\hat\omega_\eta+3-\frac{k_N^2}{k^2}\frac{\hat\omega_\eta}{\hat\omega_{\kappa}}=0.
\end{align}
Its detailed analysis is found in \SB. The stratification term $\propto k_N^2\propto N^2$ is negligible when $k\gg k_N$. In this limit, one finds $\omega_{\rm MW}=(-2\pm1)\omA^2/\om-i\eta k^2$, which describes MW damped by magnetic diffusivity. Stratification plays a role for modes with wavenumbers $k\lesssim k_N$. Buoyancy diffusion near the transition wavenumber $k\sim k_N$ where $|\omega(k_N)|\sim \omA^2/\om$ turns out to control the instability of MW. Diffusion at $k_N$ is fast ($\kappa k_N^2\gg\omA^2/\om$) or slow ($\kappa k_N^2\ll\omA^2/\om$) depending on the ratio $k_\kappa/k_N$, as seen from the identity
\begin{equation}
\label{eq:k_kappa_k_N}
 \frac{k_\kappa^4}{k_N^4}= \frac{\omA^2}{2\Omega \, \kappa k_N^2}.
\end{equation}
The solutions of the dispersion relation in the regimes of slow and fast diffusion are
\beq
    \omega_\eta=\omega+i\eta k^2
    \approx\fmw \left(-2\pm \sqrt{1+\frac{k_N^2}{k^2}}\right),
    \qquad 
    (k_\kappa\gg k_N),
\eeq
\begin{eqnarray}
\label{eq:omega_kappa_MW}
    \omega_\eta=\omega+i\eta k^2
    \approx \frac{\omA^2}{2\Omega}\times \left\{\begin{array}{lr}
\displaystyle{ -1 + i\,  \frac{k_\kappa^4}{2k^4}, } & \quad k\gg k_\kappa
\vspace*{1mm}
\\
\displaystyle{ -12\frac{k^8}{k_\kappa^8}+3i\,\frac{ k^4}{k_\kappa^4}, } &  \quad k\ll k_\kappa
     \end{array}\right\}
     \qquad 
     (k_\kappa\ll k_N).
\end{eqnarray}
One can see here that the instability $\gamma\equiv Im(\omega)>0$ appears in the regime of fast diffusion $k_\kappa \ll k_N$, and its peak growth rate $\approx\omA^2/4\Omega$ is reached at $k\approx k_\kappa$. The magnetic diffusion term $i\eta k^2$ has a damping effect on MW, as it gives a negative correction to $Im(\omega)$. The instability at $k_{\kappa}$ is not suppressed by magnetic diffusion as long as $\eta k_\kappa^2\ll\omA^2/\om$. This condition, together with $k_\kappa\ll k_N$, requires
\begin{align}
\label{eq:OmArange_singleStrat_MW}
    \left(\frac{\eta}{\kappa}\right)^{1/2}\left(\frac{ N}{\om}\right)^{1/2}\left(\frac{ \kappa k_\theta^2}{\om}\right)^{1/4}\ll \frac{\omA}{\om}\ll \left(\frac{ N}{\om}\right)^{1/2}\left(\frac{ \kappa k_\theta^2}{\om}\right)^{1/4}.
\end{align}
Note that this double inequality may be satisfied only if $\kappa\gg \eta$. Extension to multiple types of stratification below will similarly show the importance of the ratios $k_{\kappa_i}/k_{N_i}$ and $\kappa_i/\eta$.

We now turn to obtaining the MW growth rate when two types of stratification are present. We consider the case $\kapth\gg \kapmu$ relevant for stellar interiors. In the opposite case $\kapth\ll \kapmu$, the results below hold with the simple switch $\Nth\leftrightarrow \Nmu$ since buoyancy terms in the dispersion relation all have the same form. 

The growth rate $\gamma_{\rm MW}(k)$ depends on the relative order of the four stratification wavenumbers $k_{\Nth}$, $k_{\kapth}$, $k_{\Nmu}$, and $k_{\kapmu}$. One way to navigate the parameter space is to consider the effect of increasing $\omA$. As $\omA$ is increased, the condition $k_{\kappa_i}\ll k_{N_i}$ will flip to $k_{\kappa_i}\gg k_{N_i}$ since $k_\kappa$ is independent of $\omA$ while $k_N\propto \omA^{-1}$. Similar to the case of a single type of stratification described above, these flips will impact the stability of the MW modes.

For sufficiently low $\omA$ (i.e. for sufficiently weak magnetic fields), the condition of fast diffusion $k_{\kappa_i}<k_{N_i}$ is satisfied for both stratification types. The dispersion relation (\Eq~\ref{eq:MW_dispersion_normalized}) then becomes
\begin{align}\label{eq:MW_dispersion_limit1}
    D(\omega)  &=\hat\omega_\eta^2+\left(4+i\frac{k_{\kapth}^4}{k^4}+i\frac{k_{\kapmu}^4}{k^4}\right)\hat\omega_\eta+3=0.
\end{align}
Its solution $\omega_{\rm MW}(k)$ is given by \Eq~(\ref{eq:omega_kappa_MW}) with $k_\kappa^4=k_{\kapth}^4+k_{\kapmu}^4$. If $k_{\kapmu}\ll k_{\kapth}$ ($\Nmu\ll \Nth \sqrt{\kapmu/\kapth}$), thermal stratification will dominate, and the instability growth rate will peak at $k_\kappa\approx k_{\kapth}$.

More generally, whenever the condition $k_{\kapmu}\ll k_{\kapth}$ is satisfied (which is independent of $\omA$), compositional stratification weakly affects the instability of MW. Its effect remains small also when $\omA$ is increased so that $k_{\Nmu}<k_{\kapmu}$. Indeed, compositional stratification can only affect perturbations with wavenumbers $k<k_{\Nmu}$, and these wavenumbers are far below $k_{\kapth}$ (where the instability peaks), since $k_{\Nmu}<k_{\kapmu}\ll k_{\kapth}$. Thus, if $k_{\kapmu}\ll k_{\kapth}$, the instability behaves as if thermal stratification is present alone.

Next, consider the regime of $k_{\kapmu}\gg k_{\kapth}$ ($\Nmu\gg \Nth \sqrt{\kapmu/\kapth}$). A sufficiently low $\omA$ implies $k_{\kapmu}/k_{\Nmu}\ll 1$ and $k_{\kapth}/k_{\Nth}\ll 1$, so the dispersion relation is given by \Eq~(\ref{eq:MW_dispersion_limit1}), and the instability growth rate peaks at $k_\kappa=(k_{\kapth}^4+k_{\kapmu}^4)^{1/4}\approx k_{\kapmu}$, now with negligible effects of thermal stratification. However, thermal stratification can become important with increasing $\omA$, when the ratios $k_{\kapmu}/k_{\Nmu}$ and $k_{\kapth}/k_{\Nth}$ grow, and one of them exceeds unity. There are two cases:

(1) $\Nmu>\Nth \sqrt{\kapth/\kapmu}$. Then, $k_{\kapmu}/k_{\Nmu}<k_{\kapth}/k_{\Nth}$, so $k_{\kapth}/k_{\Nth}$ will exceed unity first and there will be an interval of $\omA$ where $k_{\kapmu}/k_{\Nmu}<1$ and $k_{\kapth}/k_{\Nth}>1$. In this case, thermal stratification continues to have a negligible effect on the instability, since it only affects perturbations with small wavenumbers $k<k_{\Nth}<k_{\kapth}\ll k_{\kapmu}$.

(2) $\Nmu<\Nth \sqrt{\kapth/\kapmu}$. Then, $k_{\kapmu}/k_{\Nmu}>k_{\kapth}/k_{\Nth}$, so $k_{\kapmu}/k_{\Nmu}$ exceeds unity first and there is an interval of $\omA$ where $k_{\kapmu}/k_{\Nmu}>1$ and $k_{\kapth}/k_{\Nth}<1$. In this case, the dispersion relation (\Eq~\ref{eq:MW_dispersion_normalized}) takes the form
\begin{align}\label{eq:MW_dispersion_limit2}
    D(\omega)  &=\hat\omega_\eta^2+\left(4+i\frac{k_{\kapth}^4}{k^4}\right)\hat\omega_\eta+\left(3-\frac{k_{\Nmu}^2}{k^2}\right)=0.
\end{align}
Note that $k_{\kapmu}$ no longer enters the dispersion relation, so the peak in $\gamma_{\rm MW}(k)$ at $k_{\kapmu}$ disappears. Instead, a peak around $k_{\kapth}$ can appear. The dispersion relation is a quadratic equation for $\omega_\eta$. One of its roots gives a branch of $\omega(k)$ that can be unstable, i.e. it can have $\gamma=Im[\omega(k)]>0$ for some wavenumbers $k$. This solution is approximately 
\begin{eqnarray}
\label{eq:omega_kappa_MW2}
    \omega_\eta\approx \frac{\omA^2}{2\Omega}\times \left\{\begin{array}{lr}
\displaystyle {-2 + \sqrt{1+k_{\Nmu}^2/k^2}+
i\, \frac{k_{\kapth}^4}{2k^4}\left( \frac{2}{\sqrt{1+k_{\Nmu}^2/k^2}}-1
\right), } & \quad k\gg k_{\kapth},
\vspace*{1mm}
\\
\displaystyle{ \left(3-\frac{k_{\Nmu}^2}{k^2}\right)\left(-4\frac{k^8}{k_{\kapth}^8}+i\frac{k^4}{k_{\kapth}^4}\right), } &  \quad k\ll k_{\kapth}.
     \end{array}\right.
\end{eqnarray}

An instability appears and peaks at $k\sim k_{\kapth}$ if $k_{\Nmu}<\sqrt{3}k_{\kapth}$, which sets a lower bound on $\omA$. Recalling that the dispersion relation in \Eq~(\ref{eq:MW_dispersion_limit2}) assumes $k_{\kapth}\ll k_{\Nth}$, one can see that the instability at $k_{\kapth}$ with maximum growth rate $\approx\omA^2/4\Omega$ requires $k_{\Nmu}\ll k_{\kapth}\ll k_{\Nth}$. This condition is satisfied if
\begin{align}\label{eq:kapth_instbcrit}
    \frac{\Nmu}{\Nth}\left(\frac{ \Nth}{\om}\right)^{1/2}\left(\frac{ \kapth k_\theta^2}{\om}\right)^{1/4}\ll
    \frac{\omA}{\om}\ll
    \left(\frac{ \Nth}{\om}\right)^{1/2}\left(\frac{ \kapth k_\theta^2}{\om}\right)^{1/4},
\end{align}
which requires $\Nmu\ll\Nth$. 

\bigskip

In summary:
\begin{itemize}
    \item For weak compositional stratification $\Nmu\ll \Nth \sqrt{\kapmu/\kapth}$ (equivalent to $k_{\kapmu}\ll k_{\kapth}$), $\gamma_{\rm MW}(k)$ behaves as if thermal stratification is present alone and can only have a peak at $k_{\kapth}$.
    \item For intermediate compositional stratification $\Nth \sqrt{\kapmu/\kapth}\ll\Nmu\ll \Nth $, $\gamma_{\rm MW}(k)$ can have a peak at $k_{\kapmu}$ for smaller $\omA$ and a peak at $k_{\kapth}$ for larger $\omA$.
    \item For strong compositional stratification $\Nmu\gg \Nth$,  $\gamma_{\rm MW}(k)$ behaves as if compositional stratification is present alone and can only have a peak at $k_{\kapmu}$.
\end{itemize}
Note that the growth rate can never have a peak at $k_{\kapth}$ and $k_{\kapmu}$ simultaneously. 

The intervals of $\omA$  giving the instabilities with growth rates $\approx \omA^2/4\Omega$ at $k_{\kapth}$ or $k_{\kapmu}$ may be summarized as follows.
\\
(1) Instability at $k_{\kapth}$. For weak compositional stratification $\Nmu\ll \Nth \sqrt{\kapmu/\kapth}$, thermal stratification is dominant and the interval of $\omA$ is given by \Eq~(\ref{eq:OmArange_singleStrat_MW}) with $\kappa=\kapth$ and $N=\Nth$. For intermediate compositional stratification $\Nth \sqrt{\kapmu/\kapth}\ll\Nmu\ll \Nth$, the lower bound of the interval of $\omA$ is determined by the larger of the lower bounds due to compositional stratification (\Eq~\ref{eq:kapth_instbcrit}) or magnetic diffusion (\Eq~\ref{eq:OmArange_singleStrat_MW}), since both can independently suppress instability at $k_{\kapth}$. For strong compositional stratification $\Nmu>\Nth$, the $k_{\kapth}$ mode is stable. 
\\
(2) Instability at $k_{\kapmu}$ requires a sufficient compositional stratification, $\Nmu\gg \Nth \sqrt{\kapmu/\kapth}$, and the interval of $\omA$ is given by \Eq~(\ref{eq:OmArange_singleStrat_MW}) with $\kappa=\kapmu$ and $N=\Nmu$ (combining the conditions $\eta k_{\kapmu}^2\ll \omA^2/\om$ and $k_{\kapmu}\ll k_{\Nmu}$). 

These results are summarized in Table~\ref{tab:InstabilityCriteria_thandmu} and graphically presented in Figure~\ref{fig:TwoStrat_Stab}. Note that when $\kapth,\kapmu>\eta$ the $\omA$ intervals for instabilities at $k_{\kapth}$ and $k_{\kapmu}$ merge at $\Nmu=\Nth (\kapmu/\kapth)^{1/2}$, which corresponds to $k_{\kapth}=k_{\kapmu}$. 

\section{Instability of Configurations with large gradients of $B_\phi$}
\label{ap:Large_p}

\begin{deluxetable}{cc}
\tablewidth{\linewidth}

 \tablecaption{Summary of MW instability criteria for $m=1$ modes of magnetic configurations with $p>3/2$ near the polar axis in a rotating star with $\omA\ll\Omega$. \label{tab:InstabilityCriteria_largep}}

 \tablehead{
 \colhead{Case} & \colhead{Interval of $\omA$ that gives instability with 
        $\gamma\approx \displaystyle\frac{\omA^2}{2\Omega}$}
 }
 \startdata 
    $k_{\kapth},k_{\kapmu}\ll k_\nu$ & $\displaystyle{\frac{\omA}{\om}\gg \sum_i\left(\frac{N_i}{\om}\right)^{1/2}\left(\frac{\eta k_\theta^2}{\om}\right)^{1/4} \left(1+\frac{\kappa_i}{\eta}\right)^{-1/4}}$\\[3mm]
    \hline
    $k_{\kapmu}\ll k_\nu\ll k_{\kapth}$&  $\displaystyle{\frac{\omA}{\om}\gg \left(\frac{\Nmu}{\om}\right)^{1/2}\left(\frac{\eta k_\theta^2}{\om}\right)^{1/4} \left(1+\frac{\kapmu}{\eta}\right)^{-1/4} + \frac{\Nth}{\om}\left(\frac{\nu k_\theta^2}{\om}\right)^{1/2}+\left(\frac{\Nth}{\om}\right)^{1/2}\left(\frac{\eta k_\theta^2}{\om}\right)^{1/4}}$\\[3mm]
    \hline
    $k_\nu\ll k_{\kapth},k_{\kapmu}$&  $\displaystyle{\frac{\omA}{\om}\gg \sum_i \frac{N_i}{\om}\left(\frac{\nu k_\theta^2}{\om}\right)^{1/2}+\left(\frac{N_i}{\om}\right)^{1/2}\left(\frac{\eta k_\theta^2}{\om}\right)^{1/4}}$
    \\[3mm]
 \enddata

 \tablecomments{
 The case $k_{\kapth}\ll k_\nu\ll k_{\kapmu}$ not shown in this Table is identical to that of the middle row, but with the role of the thermal and compositional stratification reversed.}
\end{deluxetable}

For rotating stars with $\Omega\gg\omA$, \SB~found that magnetic configurations with large gradients $p=\partial \ln B_\phi/\partial \ln r>3/2$ develop the TI differently than configurations with moderate gradients $p<3/2$. The configuration of magnetic fields in stars is generally unknown and such gradients may in principle occur further from the rotation axis or for more general differential rotation profiles $\Omega(R,\theta)$. Here, we generalize the results of \SB~for the case $p>3/2$ to include both thermal and compositional stratification. We will focus only on the stability analysis of MWs because the IW branch is unaffected by $p$ (see \SB), which means that results for IWs in Section \ref{ap:IWs} also hold for $p>3/2$.

For $p>3/2$, the MW branch is unstable even in ideal MHD because the large gradients of $B_\phi$ are able to overcome the stabilizing effects of rotation. In the case of a single type of stratification, MWs are unstable with a growth rate $\gamma\sim \omA^2/\om$ in an interval of wavenumbers $k_1<k<k_2$ (see \Eq~(66) in \SB). Without loss of generality, suppose this is thermal stratification. Thermal stratification suppresses the TI at wavenumbers $k<\min\{k_{\Nth},k_{\kapth}\}$ while viscosity or magnetic diffusion suppress the TI at $k>k_2=\min\{k_\nu,k_\eta (\omA/\om)\}$. Now, in the presence of compositional stratification, the instability is also suppressed at wavenumbers $k<\min\{k_{\Nmu},k_{\kapmu}\}$. Therefore, with both types of stratification present, the MWs are unstable in the interval:
\begin{align}
    k_1<k<k_2,\qquad k_1=\max\left\{\min\{k_{\Nth},k_{\kapth}\},\min\{k_{\Nmu},k_{\kapmu}\}\right\},\quad k_2=\min\{k_\nu,k_\eta \frac{\omA}{\om}\}.
\end{align} 
A fully developed instability requires $k_1\ll k_2$, which requires a minimum $\omA$. Analytic approximations for the instability criterion in different limits are presented in Table~\ref{tab:InstabilityCriteria_largep}.

\section{Microscopic diffusivities}\label{ap:Diffusivities}

We implement the microscopic diffusivities in a stellar plasma by interpolating between the non-degenerate and degenerate electrons limits, following \cite{garaud2015excitation}. The diffusivities in the non-degenerate limit may be summarized as follows.

(1) The thermal diffusivity is dominated by radiation transport,
\beq
\kapth=\frac{4ac T^3}{3\chi c_p\rho^2},
\eeq
where $c_p$ is the specific heat at constant pressure, $a$ is the radiation constant, $c$ is the speed of light, and $\chi$ is the opacity. 

(2) The magnetic diffusivity is dominated by ion-electron collisions. It is approximately given by
\beq
    \eta=\frac{\pi^{1/2}Z e^2m_e^{1/2}c^2\ln \Lambda}{16\sqrt{2}\gamma_{\rm E}(k_BT)^{3/2}}\approx 5\times 10^{11}\frac{\ln\Lambda}{(T/\mathrm{K})^{3/2}}\frac{\mathrm{cm}^2}{\mathrm{s}},
\eeq
where $e$ is the electron charge, $m_e$ is the electron mass, $Z$ is the ion charge, $k_B$ is the Boltzmann constant, $\ln \Lambda$ is the Coulomb logarithm, and $0.5\lesssim\gamma_{\rm E}\lesssim1$ is the correction for electron-electron scattering \cite{spitzer1962physics,wendell1987magnetic}. Each ion species makes a contribution to $\eta$ (and $\nu_{\rm ii}$ given below), which is weighted by its mass fraction. 

(3) The viscosity has dominant contributions from ion-ion collisions and radiation scattering,
\beq
   \nu=\nu_{\mathrm{ii}}+\nu_{\mathrm{rad}},\quad\nu_{\mathrm{ii}}=\frac{0.4m_i^{1/2}(k_BT)^{5/2}}{ Z^4e^4\rho\ln\Lambda}\approx 2\times 10^{-15}\frac{(T/\mathrm{K})^{5/2}}{(\rho/\mathrm{g}\,\mathrm{cm}^{-3})\ln\Lambda}\frac{\mathrm{cm}^2}{\mathrm{s}},\quad \nu_{\mathrm{rad}}=\frac{4aT^4}{15c\chi \rho^2},
\eeq
where $m_i$ is the ion mass \citep{braginskii1957behavior,hazlehurst1959hydrodynamics,spitzer1962physics}. 

(4) The compositional diffusivity (i.e. diffusion of ion concentration) is comparable to the ion-ion diffusion component of the viscosity $\kapmu\sim\nu_{\mathrm{ii}}$. We use the approximate expression for $\kapmu$ from \cite{michaud1993particle},
\beq
    \kappa_{\mu}=\frac{15\sqrt{2}(3+X)}{16\sqrt{5\pi}(1+X)(3+5X)(0.7+0.3X)}\frac{m_p^{1/2}(k_BT)^{5/2}}{ e^4\rho\ln\Lambda},
\eeq
where $m_p$ is the proton mass.

We now compare the magnitudes of different diffusivities. We will use the typical values ($X=0.7$, $T=2\times 10^7$\,K and $\rho= 5$\,g/cm$^3$) found in a $1.5M_\odot$ star with age $t\approx 2.852$\,Gyr (the stellar model in Section~\ref{sec:Stars}). The ratio of the compositional and magnetic diffusivities is given by 
\beq
    Cm=\frac{\kapmu}{\eta} \approx 7 \left(\frac{\rho}{5\,\mathrm{g}/\rm{cm}^{3}}\right)^{-1}\left(\frac{T}{2\times 10^7\mathrm{K}}\right)^{4}\left(\frac{\ln\Lambda}{4}\right)^{-2}.
\eeq
It can be much smaller or larger than unity, depending on local thermodynamic conditions. Note that $Cm\propto T^4\rho^{-1}$ changes rapidly near the core-envelope boundary where the density falls by orders of magnitude while the temperature decreases slowly with radius. The $Cm$ is typically larger in higher mass stars due to their higher temperatures.

The ratio of the viscosity to the magnetic diffusivity, $Pm=\nu/\eta$, is related to $Cm$ by 
\beq
    \frac{Pm}{Cm}-1\sim\frac{\nu_{\rm rad}}{\nu_{\rm ii}}\approx 8 \left(\frac{\rho}{5\,\mathrm{g}/\rm{cm}^{3}}\right)^{-1}\left(\frac{T}{2\times 10^7\mathrm{K}}\right)^{3/2}\left(\frac{\ln\Lambda}{4}\right),
\eeq
where we used $\nu\sim \kapmu+\nu_{\rm rad}$ and the Thompson scattering opacity $\chi\approx 0.2(1+X)\,\rm{cm}^2/\rm g$. The ratio $Pm/Cm$ becomes large for higher temperatures of more massive stars or lower densities at the outer edge of stellar cores (Figures \ref{fig:1p5M_proxies} and \ref{fig:1p5M_radialslice}). As a result, $Pm$ can be smaller than unity in parts of lower mass stars, but is generally much larger than unity in high mass stars. 

It is easy to verify that the thermal diffusivity is always the largest one: $\kapth\gg \nu,\kapmu,\eta$. For example,
\beq
    \frac{\kapmu}{\kapth}\approx 5\times 10^{-8} \left(\frac{\rho}{5\,\mathrm{g}/\rm{cm}^{3}}\right)\left(\frac{T}{2\times 10^7\mathrm{K}}\right)^{-1/2}\left(\frac{\ln\Lambda}{4}\right)^{-1}.
\eeq

Electrons become degenerate in the deeper interiors of evolved stellar cores. Their non-relativistic contribution to the diffusivities are implemented following \cite{hubbard1966studies}. The increased efficiency of electron conduction modifies the magnetic diffusivity, viscosity, and thermal diffusivity, while the compositional diffusivity is unaffected (ions always remain non-degenerate). The neglected relativistic effects become significant in the evolved cores of high-mass stars and can change our estimates of $Cm$ shown in Figure~\ref{fig:Cm}.
However, the relativistic corrections appear only in the region of $Cm\gg 1$ (since $\eta$ is extremely small in degenerate conditions) and do not affect the boundary of the region $Cm>1$.

\section{Influence of Differential Rotation}
\label{ap:q}
Here, we estimate the conditions for the differential rotation, $q=\partial \ln\Omega/\partial\ln R$, to be negligible in the linear stability analysis of the TI. The general dispersion relation (Equation~(A9) in \SB) has the form
\begin{equation}
  D_{q=0}(\omega,\bk)+D_q(\omega,\bk)=0, \qquad 
  D_q(\omega,k)\equiv 2q\sin\theta \frac{k_\theta k_z}{k^2} \Omega^2\left[\omega_\eta^2 -m^2\omega_A^2+2m\frac{\omega_A^2}{2\Omega}\left(\omega_\eta-\omega_\nu\right)\right]. 
\end{equation}
We focus on the magnetic configurations of primary interest, with $p=1$, and the modes with $m=1$ near the polar axis, which are most unstable. These modes have $k_z^2/k^2\approx\cos^2\theta\approx 1$.
In this paper, we investigated the TI by solving the dispersion relation $D_{q=0}(\omega,\bk)=0$, which assumes $q=0$. 
A mode $\omega(\bk)$ is expected to be weakly affected by differential rotation $q\neq 0$ if $D_q(\omega(\bk),\bk)$ is much smaller than the main terms in $D_{q=0}$ that balance to zero.

First, consider the IW modes that develop TI at $\kTI=k_\eta$ where $|\omega|\approx  2\Omega$. Then, the main terms in $D_{q=0}$ are $\mathcal{O}(\Omega^4)$. The magnitude of $D_q$ is given by
\beq
|D_q(\omega,k)|\approx 2q\sin\theta \frac{k_\theta }{k_\eta} \Omega^4,
\eeq
which is small compared to $\sim\Omega^4$ if
\beq
\label{eq:IW_q}
q \ll\frac{k_\eta}{k_\theta \sin\theta}.
\eeq

Next, consider the MW modes that develop TI at $\kTI=k_{\kapth}$, $k_{\kapmu}$, or $k_\nu$ where $|\omega|\approx \omA^2/2\Omega$. The main terms in $D_{q=0}$ are $\mathcal{O}(\omA^4)$, and the magnitude of $D_q$ is 
\beq
|D_q(\omega,k)|\approx 2q\sin\theta \frac{k_\theta }{\kTI} \omA^2\Omega^2.
\eeq
It is small compared to $\sim\omA^4$ if
\beq
\label{eq:MW_q}
q \ll\frac{\kTI}{k_\theta \sin\theta}\left(\frac{\omA}{2\Omega}\right)^2.
\eeq
The conditions for IW and MW (\Eq~(\ref{eq:IW_q}) and (\ref{eq:MW_q})) can be combined into a single expression:
\beq
q \ll\frac{\kTI}{k_\theta \sin\theta}\frac{|\omega_r|}{2\Omega}.
\eeq

\section{Suppression of Tayler instability in the compositionally stratified layer}
\label{ap:stablelayer}

\begin{figure}
    \centering
    \includegraphics[width=0.5\linewidth]{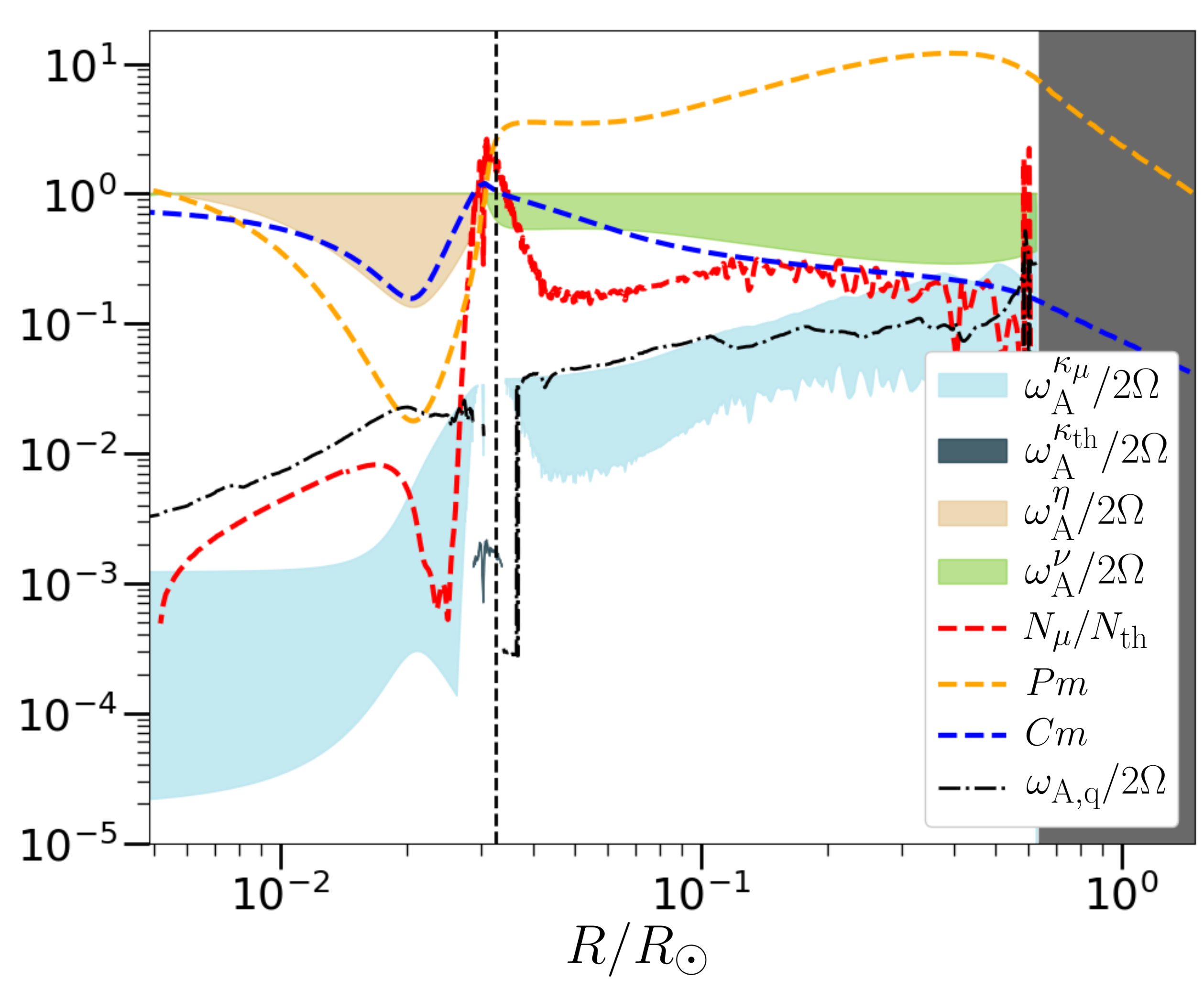}\includegraphics[width=0.5\linewidth]{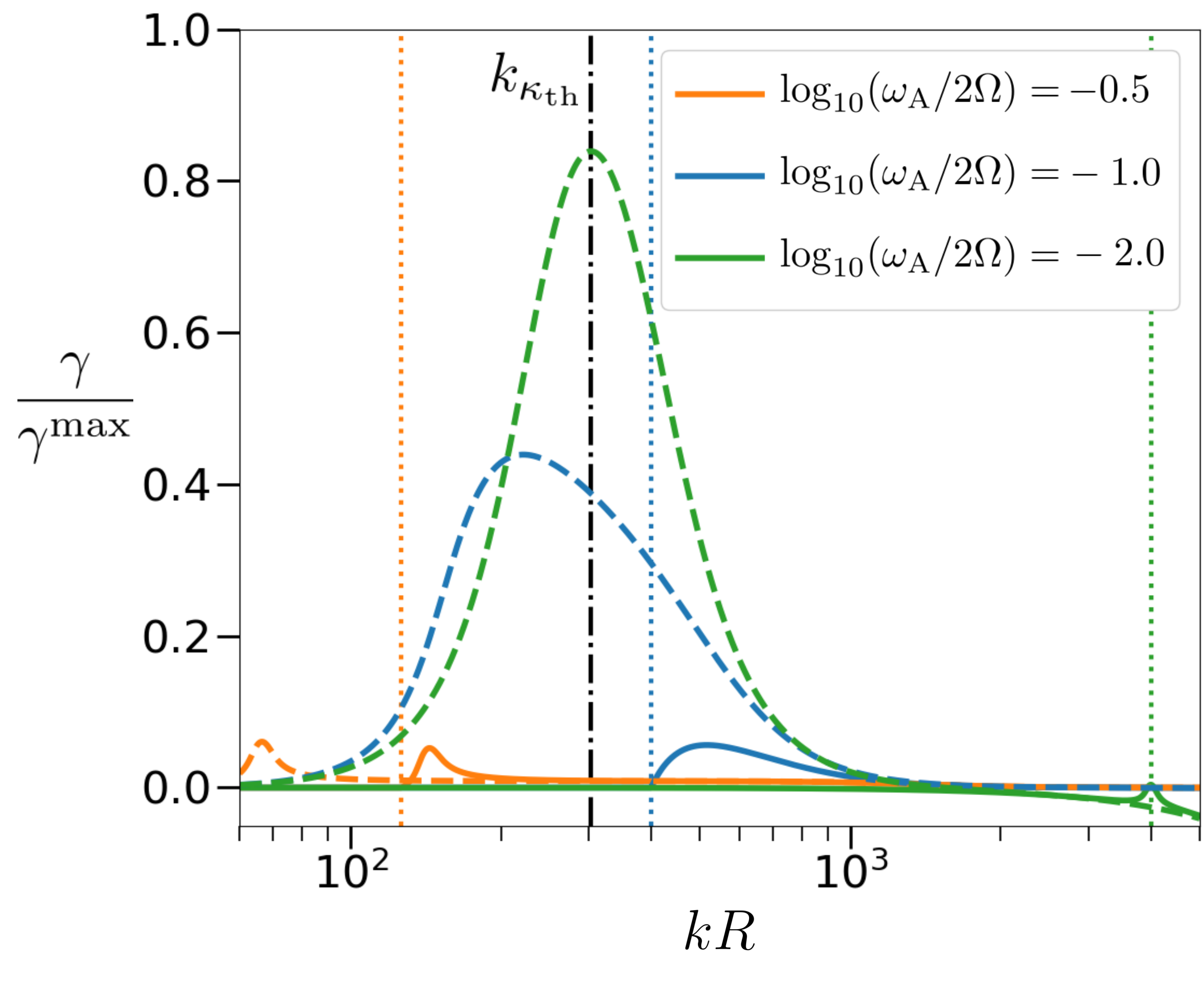}
    \caption{
    Analysis of TI modes in the $1.5M_\odot$ star with $m_{\rm He}=0.25M_\odot$ (age $t=2.8$\,Gyr), demonstrating the suppression of TI in the compositionally stratified layer above the helium core.  
    Left: Radial dependence of the instability intervals of $\omA/\Omega$ for the canonical TI modes $k_\nu$, $k_\eta$, $k_{\kapth}$, and $k_{\kapmu}$.
    Right: 
    Solid curves show the numerical solution for the one unstable MW root $\gamma(k)$ (normalized by the canonical growth rate $\gamma^{\max}=\omA^2/4\Omega$) at radius $R=0.0325R_\odot$ where $\Nmu/\Nth$ is largest; this radius is marked by the vertical dashed black line in the left panel. The other five roots of the dispersion relation are stable.  For comparison, dashed curves show $\gamma(k)$ that would be obtained with $\Nmu=0$, when only thermal stratification is present. Vertical dashed-dotted line indicates $k_{\kapth}$ (it is independent of $\omA$), and the vertical dotted colored lines indicate $k_{\Nmu}/\sqrt{3}$ for different values of $\omA/\Omega$.
    }
    \label{fig:StableLayer}
\end{figure}

When all canonical TI modes are stable according to the analytic criteria in Tables \ref{tab:InstabilityCriteria_etaandnu} and \ref{tab:InstabilityCriteria_thandmu}, the TI may still develop at non-canonical wavenumbers, with a reduced growth rate $\gamma\ll\gamma^{\max}=\omA^2/4\Omega$. As an example, we examine here the $1.5M_\odot$ star with core mass $m_{\rm He}=0.25M_\odot$ (age $t=2.8$\,Gyr) and focus on the compositionally stratified layer around the core (at radii near $0.0325R_\odot$) where $Cm\sim 1$ and $\Nmu/\Nth$ reaches its maximum $\sim 3$. Here, all the canonical modes $k_\nu$, $k_\eta$, $k_{\kapth}$, $k_{\kapmu}$ are stable or have very narrow instability strips in $\omA$ (Figure~\ref{fig:StableLayer}, left panel). The mode at $k_{\kapth}$ is stable since $\Nmu/\Nth>1$, and the mode at $k_{\kapmu}$ is stable since $Cm$ is not significantly larger than unity. The modes at $k_\eta$ and $k_\nu$ are both stable since $Pm\approx 1$, so neither $Pm\ll 1$ nor $Pm\gg 1$ is satisfied.

To study the surviving TI at non-canonical wave numbers we have numerically solved the dispersion relation at radius $R=0.0325R_\odot$. We found that one root (the MW branch with the real part $\omega_{\rm r}\approx -\omA^2/2\Omega$) gives an instability. The instability has a reduced growth rate $\gamma(k)$, reaching a peak away from the canonical wave numbers $k_{\kapmu}$ and $k_{\kapth}$  (Figure~\ref{fig:StableLayer}, right panel). The peak of $\gamma(k)$ can also be estimated analytically, as follows.

It is useful to consider first the MW growth rate $\gamma(k)$ that would occur at the same radius $R=0.0325R_\odot$ without any suppression by compositional stratification. We set $\Nmu=0$ in the dispersion relation and show the corresponding solution for $\gamma(k)$ by dashed curves in Figure~\ref{fig:StableLayer}. It has a canonical peak $\gamma\approx\gamma^{\max}$ at $k_{\kapth}$ as long as $\omA$ is in the interval $(\eta/\kapth)^{1/4}\omega_{\rm th}<\omA<\omega_{\rm th}$ (\Eqs~\ref{eq:omega_th_interval} and \ref{eq:omega_th}); this interval approximately corresponds to $10^{-4}<\omA/2\Omega<0.1$. Note that at wavenumbers $k>k_{\kapth}$, the growth rate decreases as $\gamma(k)\approx\gamma^{\max}(k_{\kapth}/k)^4$ (\Eq~\ref{eq:omega_kappa_MW2}). If $\omA/2\Omega$ exceeds $\sim 0.1$, the instability still exists but has a reduced $\gamma$.

When compositional stratification is included, the growth rate $\gamma(k)$ is shut off at wavenumbers $k<k_{\Nmu}/\sqrt{3}=k_\theta \Nmu/\omA\sqrt{3}$ (see \Eq~\ref{eq:omega_kappa_MW2}) because compositional diffusion is small ($\kapmu\gg\eta$ is not satisfied as $Cm\sim1$ in this layer). This is evident in the numerical solutions for $\gamma(k)$ with $\Nmu\neq0$ shown by solid curves in Figure~\ref{fig:StableLayer}. The cutoff wavenumber $\sim k_{\Nmu}$ is always larger than $k_{\kapth}$ in the interval $(\eta/\kapth)^{1/4}\omega_{\rm th}<\omA<\omega_{\rm th}$ due to the strong compositional stratification $\Nmu/\Nth>1$. At wavenumbers $k\gg k_{\Nmu}$, the effects of compositional stratification are small and the instability growth rate is approximately described by $\gamma(k)\approx\gamma^{\max}(k_{\kapth}/k)^4$ that is found at $\Nmu=0$. The absence of instability at $k<k_{\Nmu}/\sqrt{3}$ and the  decrease of $\gamma$ at $k\gg k_{\Nmu}$ imply that the growth rate $\gamma$ peaks at $k\sim k_{\Nmu}$ and its maximum value is $\max_k\{\gamma(k)\}\sim \gamma^{\max}(k_{\kapth}/k_{\Nmu})^4$. The suppression factor $(k_{\kapth}/k_{\Nmu})^4=(\Nth/\Nmu)^4(\omA/\omega_{\rm th})^4\ll 1$ depends on $\omA$ and is least damaging for the TI at $\omA=\omega_{\rm th}$.


\bibliography{refs}{}
\bibliographystyle{aasjournal}



\end{document}